\documentclass[pra,twocolumn,amsfonts,amssymb,amsmath,floatfix,floats,a4paper]{revtex4-2}

\usepackage{graphicx}
\usepackage{xcolor}
\usepackage{amsmath}

\begin{document}

\title{Three approaches for analyzing the counterfactuality of counterfactual protocols}

\author{Alon Wander}
\affiliation{School of Electrical Engineering, Fleischman Faculty of Engineering\\ Tel-Aviv University, Tel-Aviv 6997801, Israel}
\author{Eliahu Cohen}
\affiliation{Faculty of Engineering and the Institute of Nanotechnology and Advanced Materials\\ Bar Ilan University, Ramat Gan 5290002, Israel}
\author{Lev Vaidman}
\affiliation{Raymond and Beverly Sackler School of Physics and Astronomy\\
 Tel-Aviv University, Tel-Aviv 6997801, Israel}

\begin{abstract}
    Counterfactual communication protocols are analysed using three approaches: a classical argument, the weak trace criterion, and the Fisher information criterion. It is argued that the classical analysis leads to contradiction and should therefore be abandoned. The weak trace and Fisher information criteria are shown to agree about the degree of counterfactuality  of communication protocols involving postselection. It is argued that postselection is a necessary ingredient of counterfactual communication protocols. Coherent interaction experiments, as well as a recently introduced modification of counterfactual communication setups which eliminates the weak trace, are discussed.
         \end{abstract}
\maketitle

\section{Introduction}

Counterfactual communication of classical \cite{IFM,Penrose,Joz,Noh,Ho06,Salih} and quantum \cite{salih16,Li15,salih20,Qubit}
 information is one of the most bizarre and controversial quantum phenomena. Its  basic definition is communication without any  carriers of information  moving between Bob and Alice. However, when the carriers are quantum particles, we do not have a clear definition of their location, so the definition of quantum counterfactual communication is ambiguous. There is already a vast literature addressing the controversy surrounding counterfactual communication \cite{V07,Gisin,Li14inv,V14,V14R,count,guo15,V16,V16R,Arvid,Guo17,Cao17,Shukur,zaman18,SalRar,AV19,Hance}.
In this paper we analyze  recent arguments by Arvidsson-Shukur  and   Barnes \cite{A-SB}, henceforth A-SB, according to which postselection should not be allowed in genuine quantum counterfactual communication. We  examine various communication protocols  considered as counterfactual, as well as various approaches for defining counterfactuality. We  perform analyses using several tools,  including a powerful technical tool proposed by Arvidsson-Shukur, Gottfries  and   Barnes \cite{Shukur}, henceforth A-SGB, to evaluate counterfactuality: the Fisher information available to Alice, the recipient of the information, about a particular disturbance of the carriers of information  which occurred at Bob's site.

One might understand counterfactual communication as sending an information written in a letter without sending a letter. In fact, a more relevant model is the ancient fire (smoke) way of communication which has the form of a yes/no signal: in most protocols of quantum communication the internal degrees of freedom of the carriers of information were not used.
 
 In almost all scenarios of counterfactual communication there is a particular location in Bob's site, which we will name $B$. Bob puts or does not put an opaque object, the block, in $B$ and Alice obtains information about the presence/absence of the block in $B$, Fig~1. In all scenarios the only object which might interact with the block is a probe particle sent by Alice. A counterfactual event of communication corresponds to the situation in which one can claim that the probe particle was not present in $B$. 
 

\begin{figure}
  \includegraphics[width=0.9\linewidth]{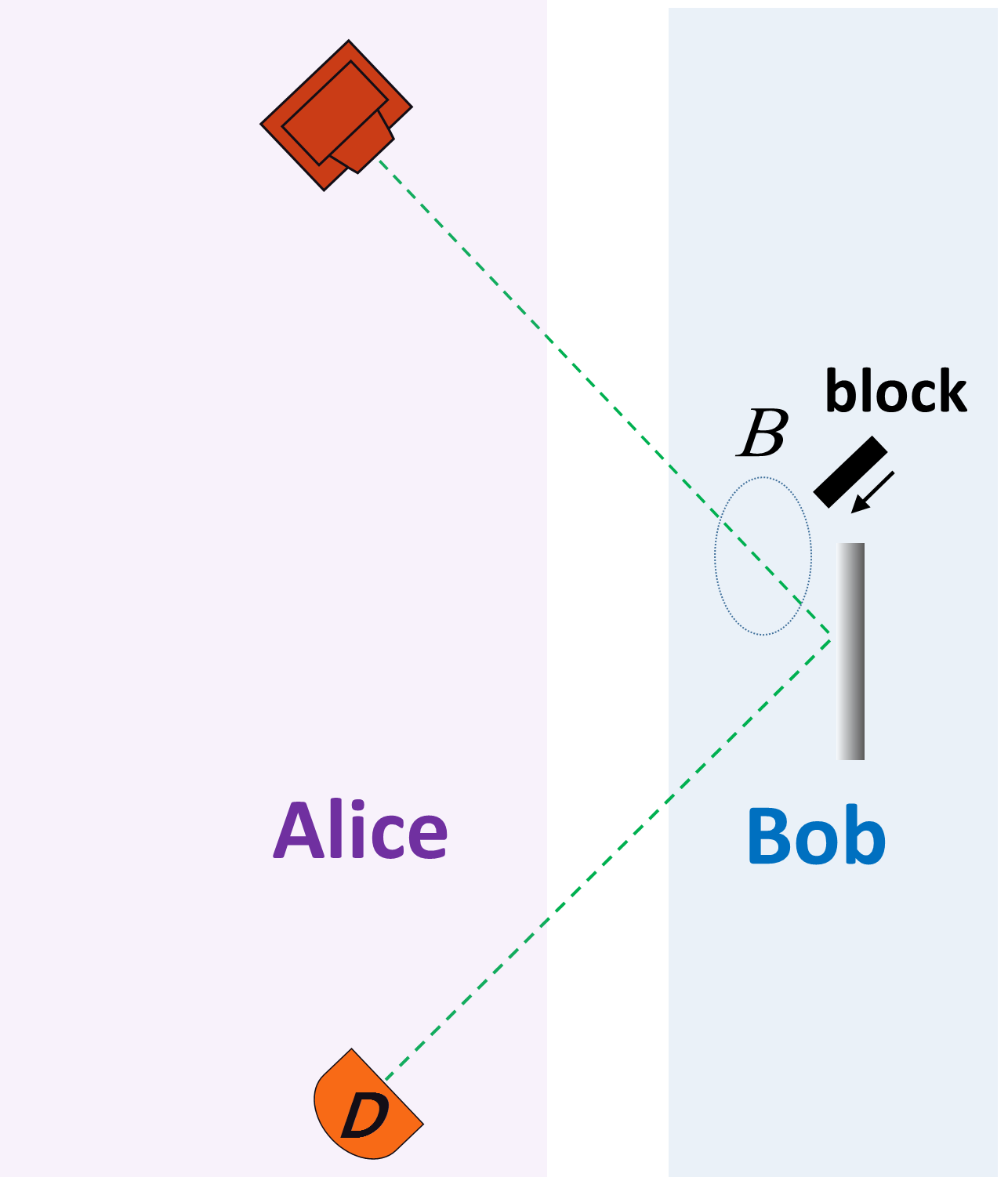}
  \caption{{\bf A simple communication protocol with a probe particle.} Alice sends a probe particle at a known time and Bob blocks (for bit 1) or does not block (for bit 0) the path by an opaque object placed in $B$ which is located in his site.}
  \label{fig:setup}
\end{figure}

We will distinguish several communication tasks to be performed in a counterfactual way:

(i) Alice finds that the block is in $B$.

(ii) Alice finds that the place $B$ is empty.

(iii) Alice finds what is the situation in $B$, i.e. whether it is empty or it is occupied by the block.

We will evaluate counterfactuality according to three possible criteria:

(a) Naive classical argument: the probe was not in $B$ because it could not reach Alice's detector while moving through $B$.

(b) Weak trace: the probe was not in $B$ because it left a trace in $B$ which is much smaller than the trace of a localized probe passing through $B$.

(c) Fisher information: the probe was not in $B$ because the Fisher information about the small disturbance in $B$ carried by the probe reaching Alice is much smaller than the Fisher information of a localized probe passing through $B$.

For each protocol it will be defined which events in Alice's site are legitimate according to the protocol. 
We will consider the question of counterfactuality only for legitimate events. Note that these legitimate events might not  necessarily correspond to successful communication. In some protocols an error might happen even with ideal devices. The error in communication does not prevent us from considering the event in the evaluation of the counterfactuality of the protocol. All legitimate events have to be taken into account.

The structure of the paper is as follows. In Section II we discuss the crucial role of postselection in counterfactual protocols. In Sections III-V we analyze the counterfactuality of several protocols: In Section III according to the classical criterion,  in Section IV according to the weak trace criterion, and in Section V according to the Fisher information criterion. In Section VI we analyze the difference between coherent and incoherent interaction of the probe according to the weak trace and Fisher information criteria. Section VII describes a recently proposed modification of counterfactual communication protocols which essentially removes the weak trace of the probe (and reduces Alice's accessible Fisher information). Section VIII summarizes the results of our analysis.

\section{Postselection}

A-SB \cite{A-SB} suggested that   genuine quantum counterfactual communication should not include postselection. As will be discussed below, we find this proposal unacceptable. Counterfactual communication without postselection, if possible, contradicts our basic understanding about nature. While the current counterfactual protocols are surprising, a counterfactual protocol without postselection is impossible as it essentially describes an action at a distance. Today, the only (and indeed spooky) action at distance in modern physical theories appears in the collapse process as part of quantum measurements. The postselection of a particular result corresponds to this collapse (or effective collapse if we adopt the many-worlds interpretation \cite{Everett} or de Broglie-Bohm interpretation \cite{Bohm}). Without postselection, we have a unitary process which respects  causality. The influence of Bob's action of placing (or not placing) the object in $B$  propagates in a continuous way to Alice. Whatever propagates represents the probe particle, and therefore a fully counterfactual communication is impossible unless postselection is employed.

Historically,  the term ``counterfactual'' was introduced by Penrose \cite{Penrose} for describing  the Elitzur-Vaidman (EV) interaction-free measurement \cite{IFM} (IFM), where the postselection plaid a crucial role. This is the above task (i), Alice finds that an opaque object is in $B$ without any probe particle ``being'' in this place. The object can be a ``bomb'' which explodes when any single particle reaches it, but still it can be found without explosion. The EV device is a  Mach-Zehnder interferometer (MZI)  with one arm passing through point $B$ tuned in such a way that  one port is 100\% dark, Fig.~2a. The wave packet of the probe particle splits into two parts, and the probe can reach point $B$. According to the protocol, the only legitimate (postselection) event is  when Alice finds the probe particle in the dark port $D$, Fig.~2b. It is possible only when the object is present in $B$, because otherwise she could not find the probe particle. Alice can  also claim that the probe particle was not in $B$, because it could not pass through the block to Alice.

\begin{figure}
  \includegraphics[width=0.9\linewidth]{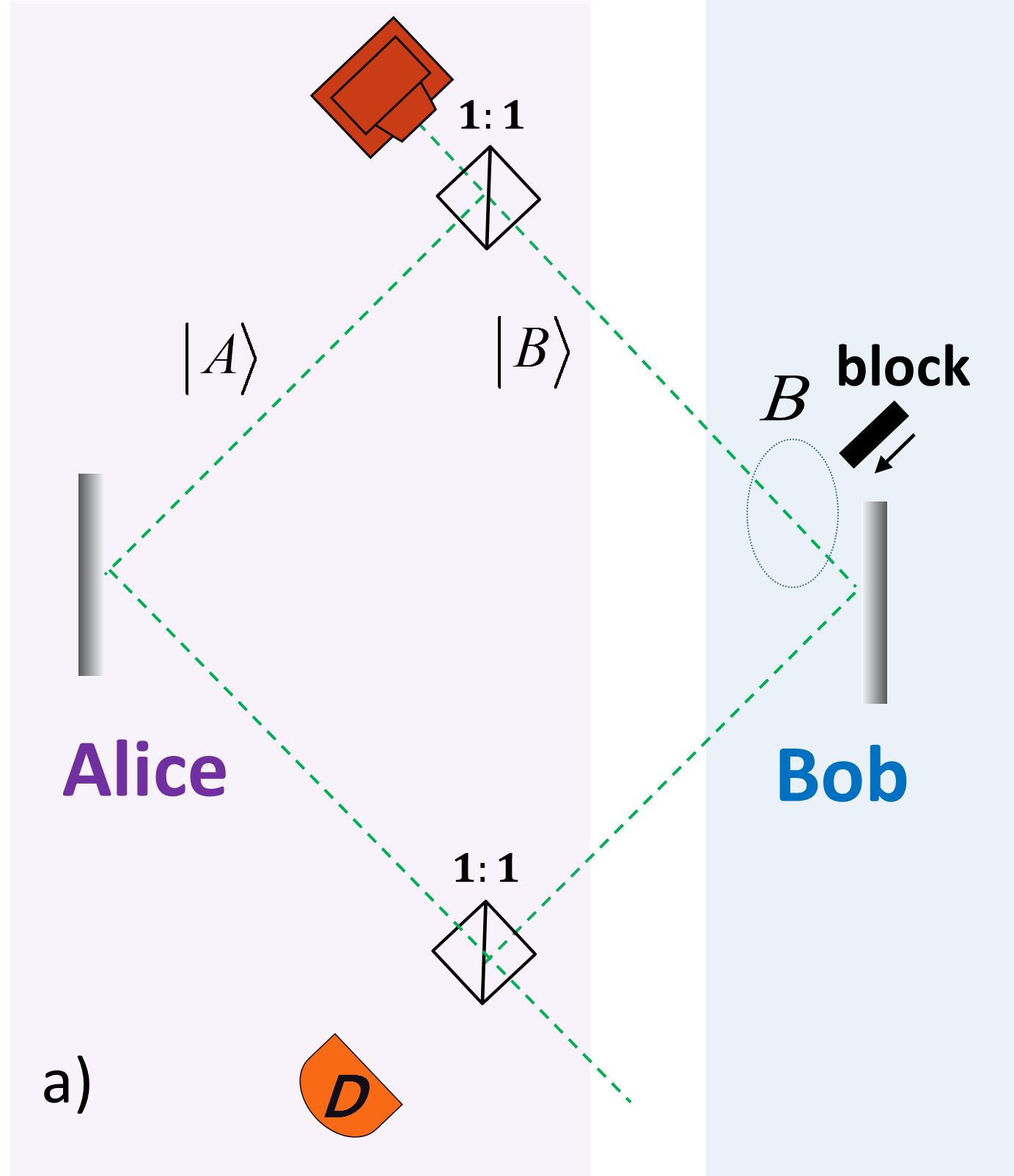}
  
  \includegraphics[width=0.9\linewidth]{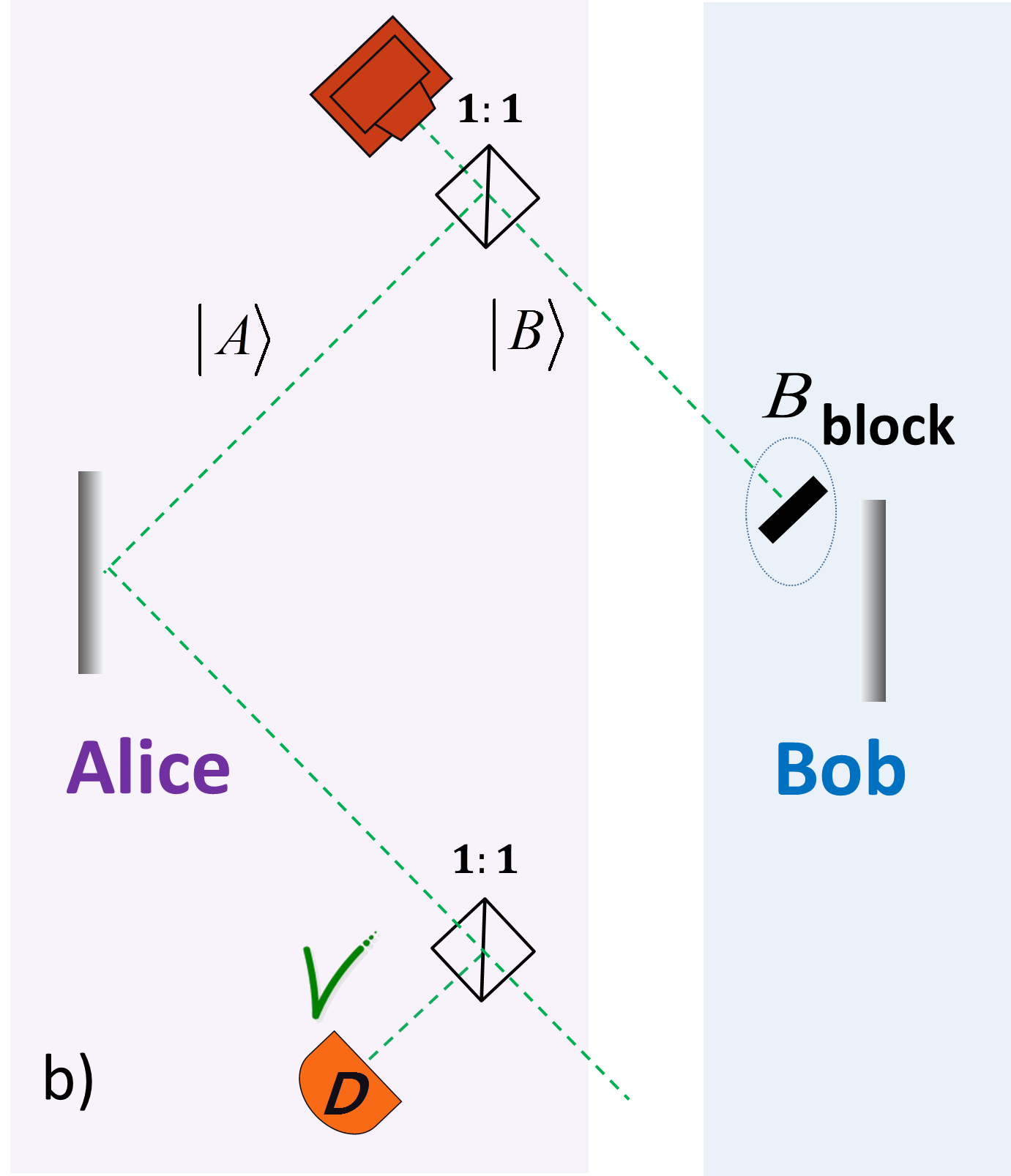}
  \caption{{\bf Counterfactual measurements of the presence of a block. } a) The MZI is tuned such that the port with detector $D$ is dark due to destructive interference. b) A click of the detector can happen only if the block is present in $B$. }
  \label{fig:setup}
\end{figure}

The probe particle could have been in $B$, but actually it was not. This is the reason for the term ``counterfactual''. The possibility of the probe particle being there was enough for obtaining information about $B$. The quantum state could collapse to $B$, but it did not. Finding the probe in Alice's detector amounts to the collapse of the quantum state of the probe to the other arm. In the many-worlds interpretation, the probe was in $B$ within a parallel world, while in Bohmian mechanics the empty wave of the probe was in $B$. In the EV device we consider postselection on the detection of the dark port, only then we can claim that the probe was not in $B$.

There was a  controversy about the claim that the probe was not near the bomb in the EV IFM \cite{EVmeaning}. The wave of the probe, before the collapse, reached the bomb. What made this quantum counterfactual method acceptable as a significant achievement (one of the seven wonders  of quantum world \cite{NewSci}) was that it performed a task which could not be accomplished in the framework of classical physics. Alice sometimes gets definite information about the presence of an object in $B$ without leaving any trace in $B$. 

Note, that it is trivially possible even in classical world if there is prior information of the following kind: the object is either in $B$ or in place $A$ near Alice. Alice looks at $A$ and if she finds nothing, she knows with certainty that the object is in $B$ without leaving any trace there. The object might be a bomb exploding whenever a probe reaches it, Alice finds its presence in $B$ without explosion. In contrast, the quantum method achieves this without prior information.

Let us describe  a classical counterfactual protocol with postselection similar to the one considered by A-SB \cite{A-SB}.  Alice sends a probe particle every minute. Every minute Bob chooses a  bit by putting   or not putting   an opaque object in $B$, as in Fig.~1. But instead of the standard agreement, i.e. putting the block in $B$ for 1, and leaving it out for 0, on every odd minute  putting   the block by Bob corresponds to bit 1  and not  putting to bit 0, while on every even minute the rule is the opposite, blocking for 0 and leaving open for 1.  Alice and Bob agree to perform postselection by considering only events when Alice does not receive the probe particle.

Although technically, in all postselected cases, Alice  knows  Bob's bit without any probe particle reaching her, there is nothing counterfactual here. There is no other option  which could have happened, but did not actually happen. If Bob's object is a bomb, Alice could not find it without explosion. 

The  task that can be achieved using the quantum EV method, whose counterfactuality is uncontroversial, is quantum key distribution \cite{Noh}. If we consider a passive eavesdropper Eve, who only observes the transmission channel after the transmission, then it is absolutely secure. In this protocol, there are two copies of the EV system as in Fig.~2, named 0 and 1. Bob randomly blocks one of them and Alice randomly chooses  one of them to send her probe particle. She announces every time she gets a click in the dark output port of one of the EV interferometers. These are the events of legitimate postselection of the protocol. In all these cases Alice and Bob choose the same device (the necessary requirement for the dark port count). This way, Alice and Bob obtain  a common random (secret) bit. Indeed, a passive Eve cannot see  any difference between the devices: no probe particle is present in any of the transmission channels. The footprints of the particles Eve might  observe in other, not postselected, runs provide no information about the generated common key. This is in clear contrast to the A-SB communication protocol described above. A passive Eve, observing the footprints of the probe in the channel obtains information about its presence and absence and it provides information about Bob's bit.

Postselection plays a crucial role in the EV original protocol. The probability of success of a single application of the protocol is $\frac{1}{4}$ and in half of the cases the bomb explodes. The modification of the protocol employing the quantum Zeno effect \cite{IFMKwiat} allows one to make the probability of success arbitrarily close to 1 with the price of increasing the duration of the protocol and creating numerous different possibilities of failure in which the bomb explodes at different times. Given  the click in the dark port, which now is very probable, the protocol is clearly counterfactual. Without postselection, we have numerous failure possibilities of counterfactuality. Although the probability of all these (non-counterfactual) cases together is vanishingly small, they do play a crucial role in the procedure, so it is hard to call the protocol (without removing them via postselection) counterfactual.



\section{Classical naive counterfactuality}

Counterfactual protocols succeed because they provide  a possibility for a probe to be in Bob's site $B$, but in the actual, postselected case, the probe was not there. How shall we decide that the quantum probe, which has no clear definition for its location in the past, was not in $B$? The first approach is based on a classical picture. It tacitly assumes that at every moment the particle is in a particular place, even if we might not know it. The classical argument seems very innocent: if the particle could not have been in $B$, then it was not in $B$.

In the EV IFM, Fig.~ 2., the block can be  a sensitive bomb, so  the presence of the probe in $B$ leads to an explosion. Since there was no explosion, the probe was not in $B$, or if the block is just an opaque object, the presence of the probe in $B$ leads to its absorption. Since the probe was found in Alice's detector, it was not absorbed, therefore the probe was not in $B$.

For the IFM telling us that a particular place $B$ at Bob's site is empty, the argument looks very similar. The  nested MZI is tuned in such a way that the photon entering
the inner interferometer leaves the external interferometer without reaching its final beam splitter, Fig.~3a. It is also tuned such that if there is a block in $B$, Alice's detector $D$ cannot click, Fig.~3b.  Thus, if the probe reaches  detector $D$, the inner interferometer must be empty. But if the probe was in the undisturbed inner interferometer it could not reach Alice's detector. Therefore, the probe detected by $D$, could not have been in $B$.

\begin{figure}
  \includegraphics[width=0.95\linewidth]{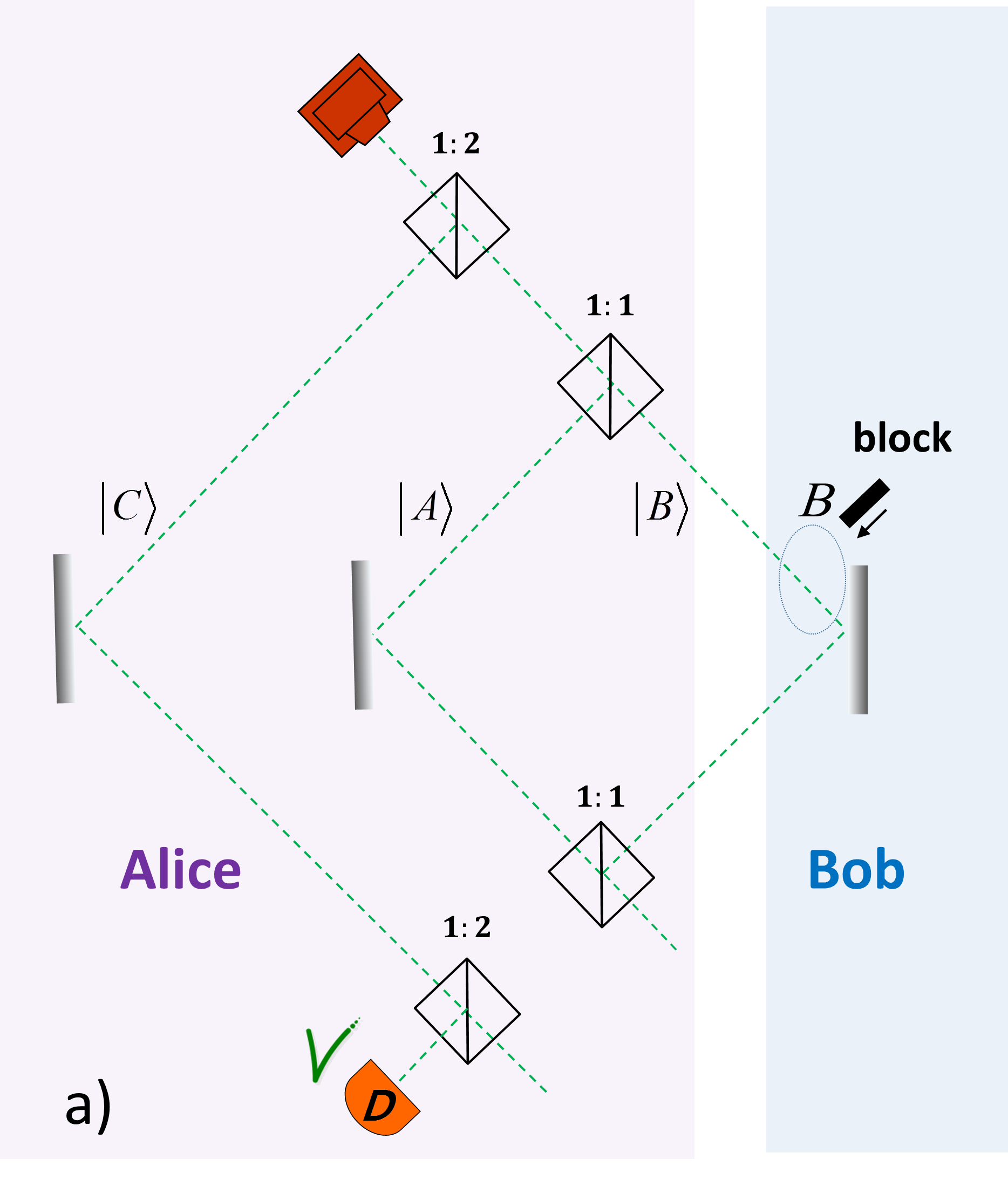}
  \includegraphics[width=0.95\linewidth]{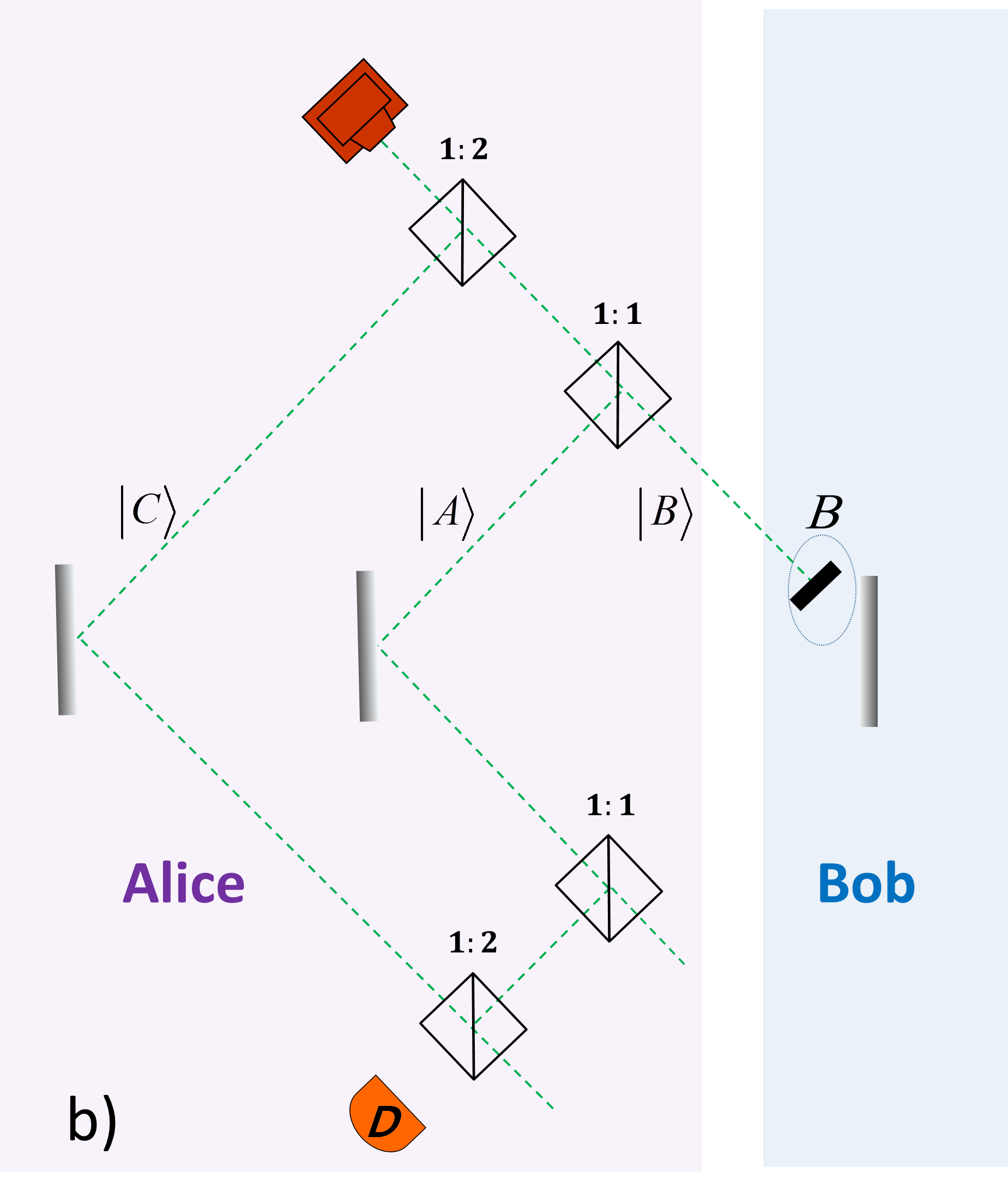}
  \caption{{\bf Counterfactual measurements of the absence  of a block. }  a) A small MZI, nested in the right arm of the large MZI, is tuned such that  the probe  leaves the large interferometer without reaching its final beam splitter. b) The large interferometer is tuned such that when $B$ is blocked, the port of detector $D$ is dark due to destructive interference.
 A click at the  detector can happen only if the block is absent in $B$. }
 \end{figure}

Allowing postselection, we can construct a counterfactual communication protocol which distinguishes both cases: presence and absence of the object in $B$, see Fig.~4. (The protocol was proposed earlier  \cite{V2019}, but there was an error in Fig.~1 therein. The   beam splitter $BS_1$ must have  reflectively 3:32, and not 3:8 as appeared there.)  Detection in one of the detectors, $D_0$ or $D_1$, tells us without an error if the object is present or absent in $B$. This method has a small probability of success, in most cases none of these detectors clicks, so it is hardly useful for  practical applications, but it is important conceptually for the rare cases when  Alice detects  the first photon  she sends.

 If we allow a small chance for an error, then (apart from significant experimental difficulties) one can construct an efficient counterfactual communication protocol applying the quantum Zeno effect \cite{Ho06,Salih,salih16,salih20,Li15}, see Fig. 4. These protocols might have an error regarding the transmitted bit, while counterfactuality, if we consider our naive argument and ideal devices, remains precise. In these protocols there are many paths and they are all  blocked or all free. In case they are blocked, the particle could not be at Bob's place because then it is certainly absorbed by one of the blocks, and if they are free, then again, particles going into the inner interferometers with parts at Bob's site cannot come back to Alice due to destructive interference.

\begin{figure*}
  \includegraphics[width=14cm,height=21cm]{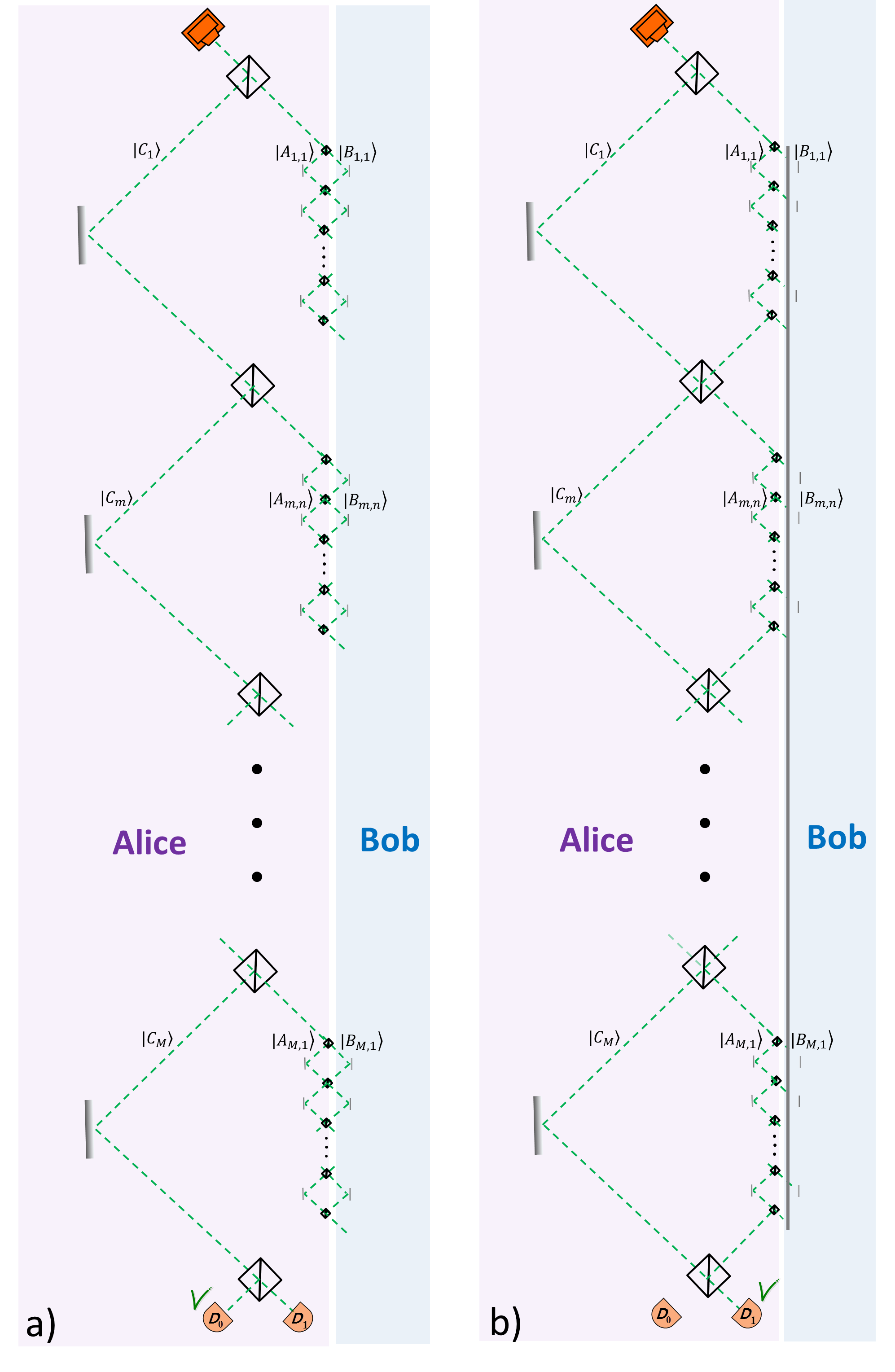}
  \caption{{\bf Counterfactual communication employing the quantum Zeno effect. }   a) When the small interferometers are empty, the probe entering the right arm of any external interferometer  does not continue inside the interferometer, making the probability of detection by $D_1$ very small.   b) When all the right arms of the small interferometers are blocked,  there is destructive interference toward detector $D_0$. }
\end{figure*}


Although the argument for counterfactuality of the IFM of the absence of an opaque object  in $B$, Fig.~3, sounds as persuasive as the argument for the IFM of the presence in $B$ (Fig.~2) it leads to a paradox which shows that it cannot hold. Let us see how the paradox arises.

We consider a successful protocol of finding that $B$ is empty when the probe is detected at $D$.  The probe moved from the source to the detector and at the intermediate time it could have been in $A$, $B$ and $C$. Since, by our argument it could not have been in the inner interferometer it must have been in $C$ (see \cite{past,Berge,PV,PVrep}).

In standard quantum mechanics, systems are described by their quantum states. Any question about time $t$, including the location of the particle, should be answered from knowledge of the quantum state at time $t$. In our case we need to add information about postselection, i.e. detection by detector $D$. This is equivalent to a measurement of a particular state at time $t$, which is calculated through unitary evolution backward in time from the postselection at $D$. So, the forward-evolving state at time $t$ specified by the preselection  and the backward-evolving state at that time specified by the  postselection  provide complete information about the probe particle.
If the question of where the particle was located at time $t$ makes sense, it has to be specified by these states. 

Our naive classical arguments, which included information about history and future relative to time $t$ made us believe that the probe particle is in $C$ and was not in $B$. The paradox is that there is a symmetry between $B$ and $C$ both for preselected state $|\Psi \rangle$  and the postselected state $\langle\Phi|$ at time $t$. In fact, these are the states of the 3-box paradox \cite{AV91}, \begin{equation} \label{3box}
\begin{split}
|\Psi \rangle= \frac{1}{\sqrt3}(|C\rangle+|B\rangle+|A\rangle),\\
\langle\Phi|=\frac{1}{\sqrt3}(\langle C|+\langle B|-\langle A|).
\end{split}
\end{equation}
The two-state vector $\langle\Phi|~|\Psi \rangle$ is the concept of  of the  two-state vector formalism (TSVF) \cite{AV90}.   Standard quantum formalism and the TSVF are equivalent, but we use the latter because it is better suited for analyzing pre- and postselected quantum systems.
(The Consistent Histories approach \cite{Griff} leads to an interesting alternative analysis of such situations  \cite{GriffCo,GriffCoV,GriffCoVR,GriffCoS,GriffCoSR}, but this  goes beyond the scope of the current work.)

If we consider the protocol for IFM of the presence of the object, we also have the symmetry between the arms $A$ and $B$ in the pre- and postselected states:
\begin{equation} \label{IFM}
\begin{split}
|\Psi \rangle= \frac{1}{\sqrt2}(|B\rangle+|A\rangle),\\
\langle\Phi|=\frac{1}{\sqrt2}(\langle B|-\langle A|).
\end{split}
\end{equation}
 (The ``-'' sign in the postselected state can also belong to $|A\rangle $ since an overall phase is of no importance.) However, this symmetry does not exist at any intermediate time. While in the case of the absence of the object nothing disturbs the forward- and backward-evolving states Eq. (\ref{3box}), the opaque object blocks the path $B$, so  before the opaque object the two-state vector description is:
\begin{align} \label{IFMbefore}
|\Psi \rangle= \frac{1}{\sqrt2}(|B\rangle+|A\rangle),\\
\langle\Phi|=\langle A|,~~~~~~~~~~~~~~~
\end{align}
 and at the time the wave packet of the probe passes the object,
the two-state vector description is:
\begin{equation} \label{IFMafter}
\begin{split}
|\Psi \rangle= |A\rangle,~~~~~~~~~~~~~~\\
\langle\Phi|=\frac{1}{\sqrt2}(\langle B|-\langle A|).
\end{split}
\end{equation}
At no moment of time do we have symmetry between $A$ and $B$: the probe particle was not in $B$ at any intermediate time.

\section{Weak trace counterfactuality}

The classical naive approach to counterfactuality leads to a consistency paradox as explained above, so it cannot be accepted. We can just refrain from discussing the location of the quantum probe as Bohr preached or we should look for another approach. We will discuss now the weak trace approach \cite{past}, according to which the quantum particle was where it left a trace of the order of the trace left by a well-localized particle being there.

We are following  a model  in which  the state of the particle  passing through a channel is not changed, but the quantum state of the environment in $B$,  originally described by  $| \chi \rangle$, is modified due to the passage  of the localized particle \cite{PNAS}:
\begin{equation}\label{eq::iaSingle}
|\chi \rangle \rightarrow|\chi^\prime \rangle \equiv \sqrt{1-\epsilon^2} |\chi \rangle + \epsilon |\chi^\perp \rangle ,
\end{equation}
where $| \chi^\perp \rangle$ denotes the component of $| \chi^\prime \rangle$ which is orthogonal to $| \chi \rangle$ and its phase is chosen such that $\epsilon > 0$.

When the particle is not a localized wave packet, but is in a superposition of several spatial locations, which is later detected in a state corresponding to another superposition,
the local environment is changed in a similar way. The environment obtains the same orthogonal component due to the interaction and the only difference is the amplitude of this component. In the TSVF, the modified amplitude is given by a very simple expression. The modification (for    $\epsilon \ll 1$) is \cite{PNAS}:
\begin{equation}\label{eq::chi''}
|\chi \rangle \rightarrow |\chi'' \rangle = \sqrt{1-|({\rm \bf P}_B)_w ~ \epsilon~|^2} ~|  \chi \rangle +  \left({\rm \bf P}_B\right)_w ~ \epsilon ~ |\chi^\perp \rangle ,
\end{equation}
where $\left({\rm \bf P}_B\right)_w$  is the weak value of the projection on $B$  for the forward-evolving (preselected) state $| \psi \rangle$ and the backward-evolving (postselected) state  $\langle \phi |$
\begin{align}\label{eq::wvIdealMZI}
\left( {\rm \bf P}_B \right)_w &\equiv \frac{ \langle \phi | {\rm \bf P}_B | \psi \rangle }{\langle \phi | \psi \rangle} .
\end{align}
In this case, the probability to find an orthogonal component after the end of the protocol is 
\begin{align}\label{probweak}
{\rm Prob} =|({\rm \bf P}_B)_w ~ \epsilon~|^2 .
\end{align}

An example of a trace left by a particle on a channel's environment is the change of the state of mirror $B$ in the MZI,  see Fig.~5. If the path $B$ is blocked, Fig.~5a, the quantum state of the mirror   $| \chi  \rangle$ remains unchanged. If the other arm is blocked, Fig.~5b, the particle certainly bounces off the mirror $B$ and after the particle is detected, its  state is  $| \chi'  \rangle$. Finally, if both arms are open, Fig.~5c, the state of the mirror $B$ (after detection of the particle) is  $| \chi''  \rangle$.

The weak value provides a simple expression for the weak trace criterion. In a single weak coupling of a localized wave packet, the probability to find an orthogonal component in the environment is  $\epsilon^2$ and in the pre- and postselected situation it is multiplied by $|({\rm \bf P}_B)_w |^2 $. Hence, if $|({\rm \bf P}_B)_w | $ is of order 1 (or more), then the particle was in $B$.

According to the weak trace criterion, the EV IFM, Fig.~2, is fully counterfactual. After a successful event of Alice detecting the click at $D_0$, there is no orthogonal component of the environment near the opaque object. In the TSVF it can be easily seen: there is no overlap of the forward- and backward-evolving wave functions in $B$, therefore $\left({\rm \bf P}_B\right)_w=0$,  and together with it, the amplitude of the orthogonal component vanishes.

On the other hand, the simple IFM of the absence of the object, Fig.~3 is not counterfactual. We consider postselection only on the click of detector $D_0$, the only legitimate scenario of finding the absence of the object. Then, $\left({\rm \bf P}_B\right)_w=1$, \cite{V07}  and the trace is the same as in a non-counterfactual communication with one probe particle. The weak trace criterion for counterfactuality does not lead to the contradiction we have seen above in the context of the naive classical argument for counterfactuality. The symmetry of the quantum description between $B$ and $C$ is not broken by the weak trace in the two arms, $\left({\rm \bf P}_B\right)_w=\left({\rm \bf P}_C\right)_w$. 
Note, however, the controversy about the particle path through such nested MZI and the trace the particle leaves \cite{pra88,pra88R}.

\begin{figure}
  \includegraphics[width=0.7\linewidth]{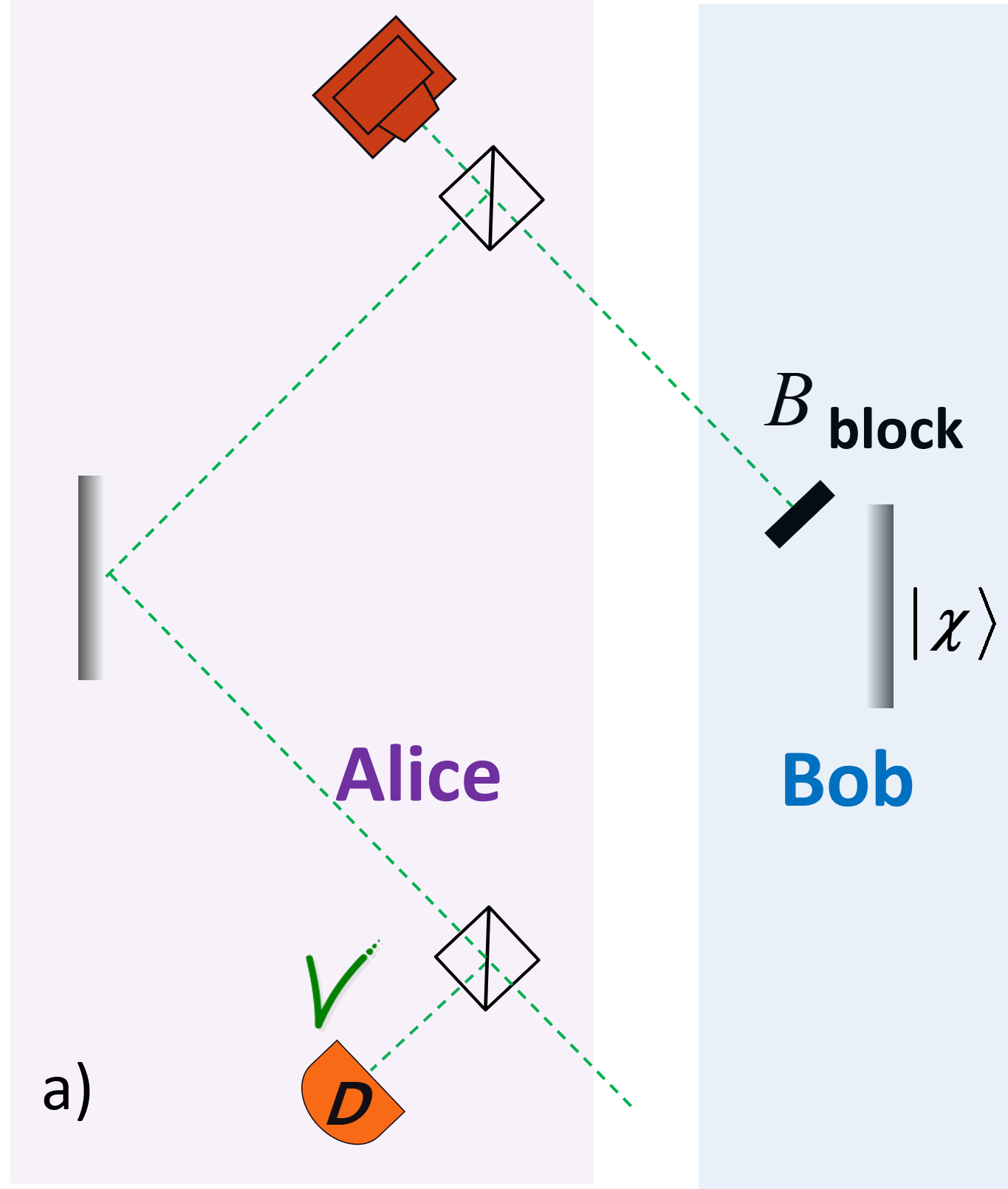}
  
  \includegraphics[width=0.7\linewidth]{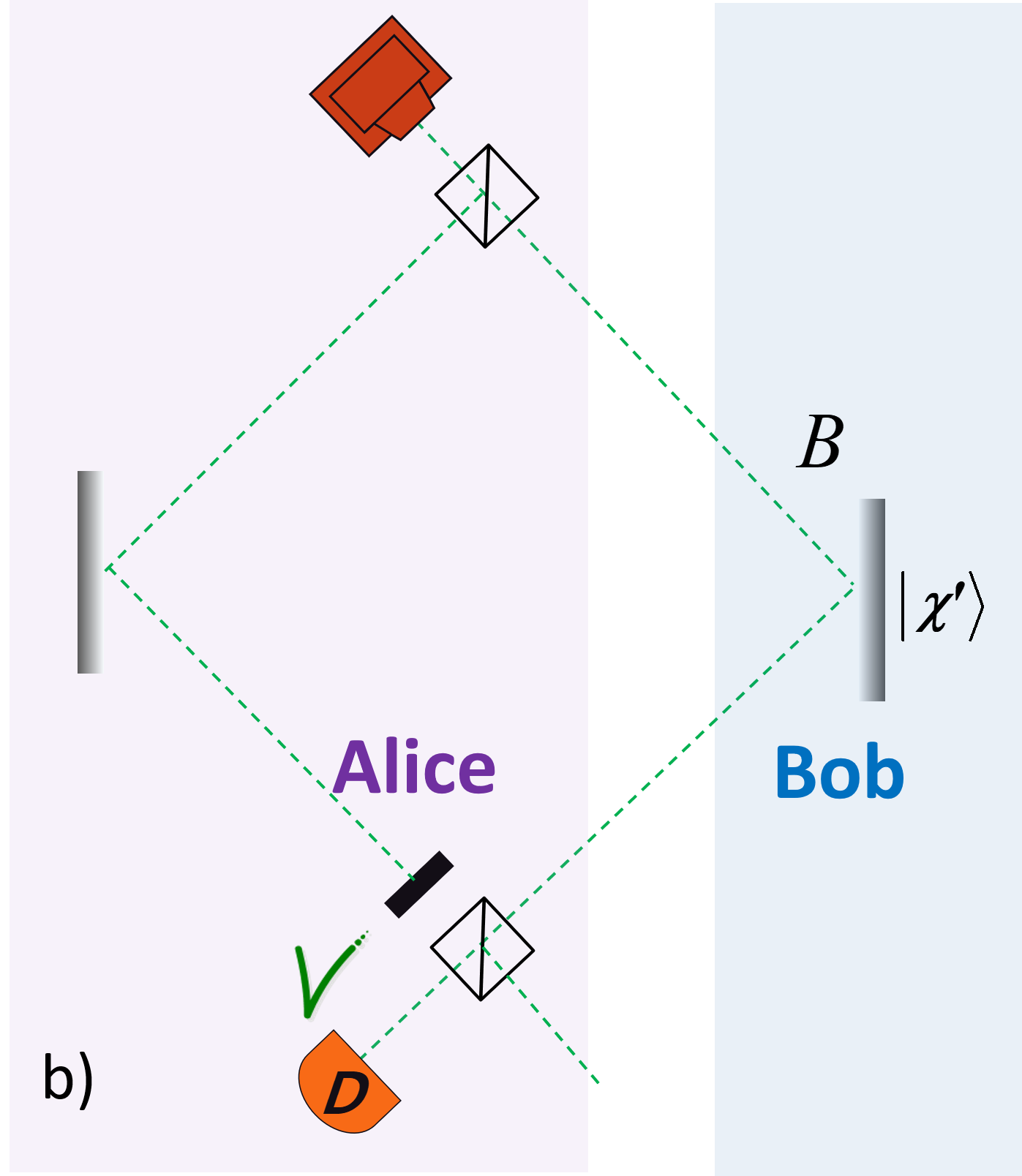} 
  
  \includegraphics[width=0.7\linewidth]{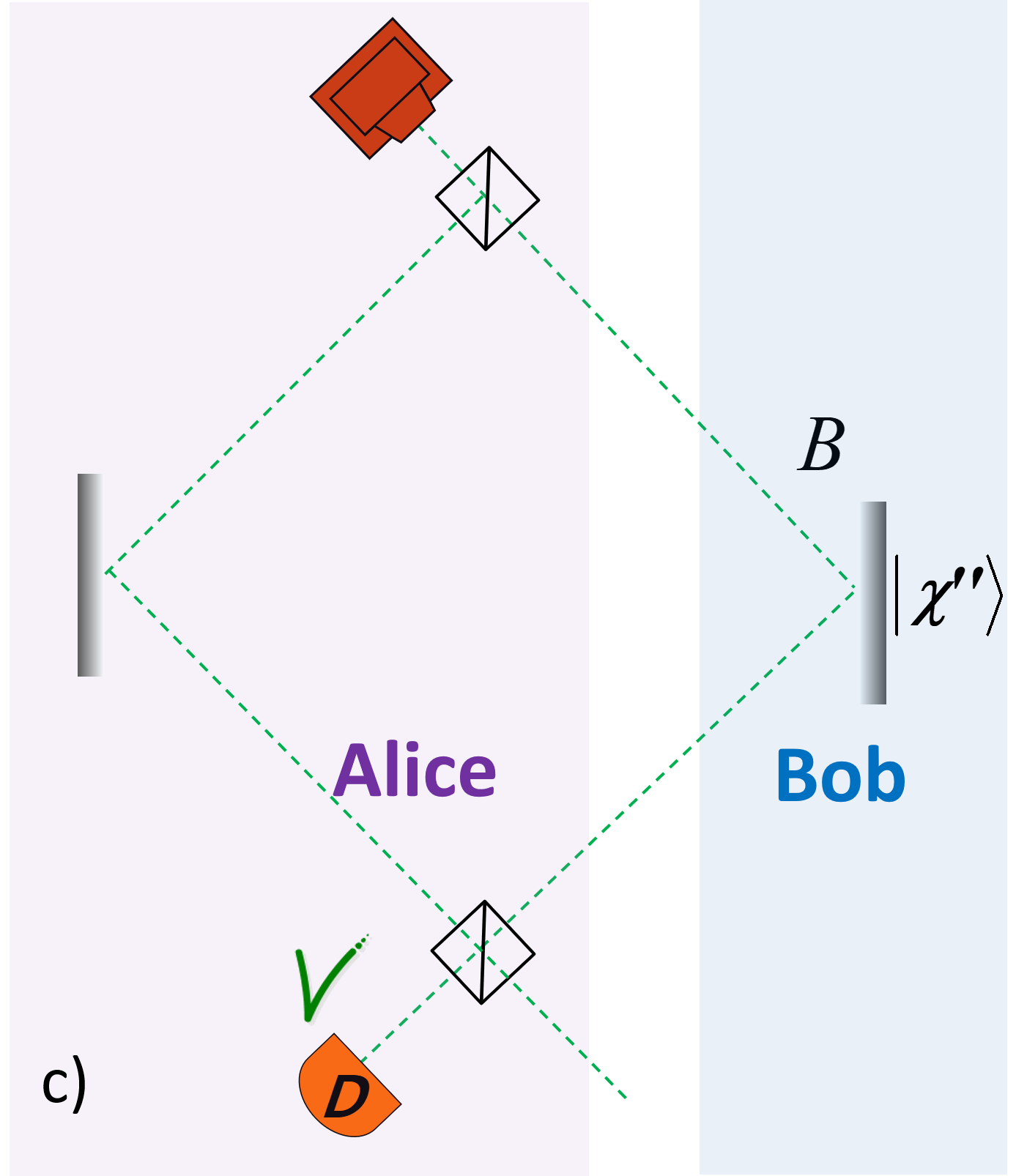}
  \caption{{\bf The weak trace left on a mirror in the MZI.}   a) When the arm of the interferometer is blocked, the quantum state  of the mirror in this arm, $| \chi  \rangle$, is not changed. b) When the other arm  is blocked, the particle, which has only one open arm, changes the mirror state according to Eq. ($\ref{eq::iaSingle}$). c) When both arms are open, the final state of the  mirror is given by Eq. ($\ref{eq::chi''}$).}
 \end{figure}

\begin{figure}
  \includegraphics[width=0.95\linewidth]{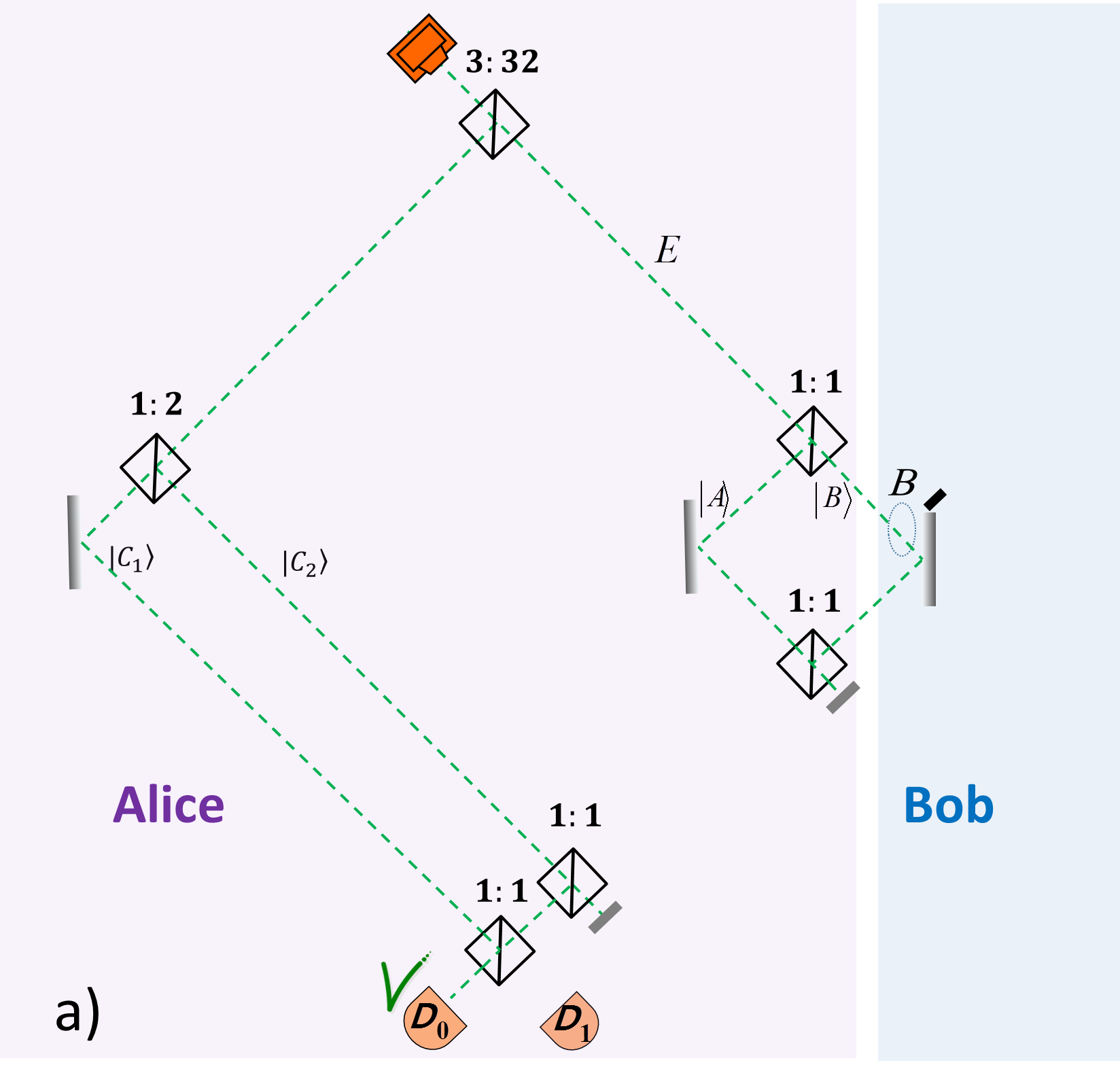}
  
  \includegraphics[width=0.95\linewidth]{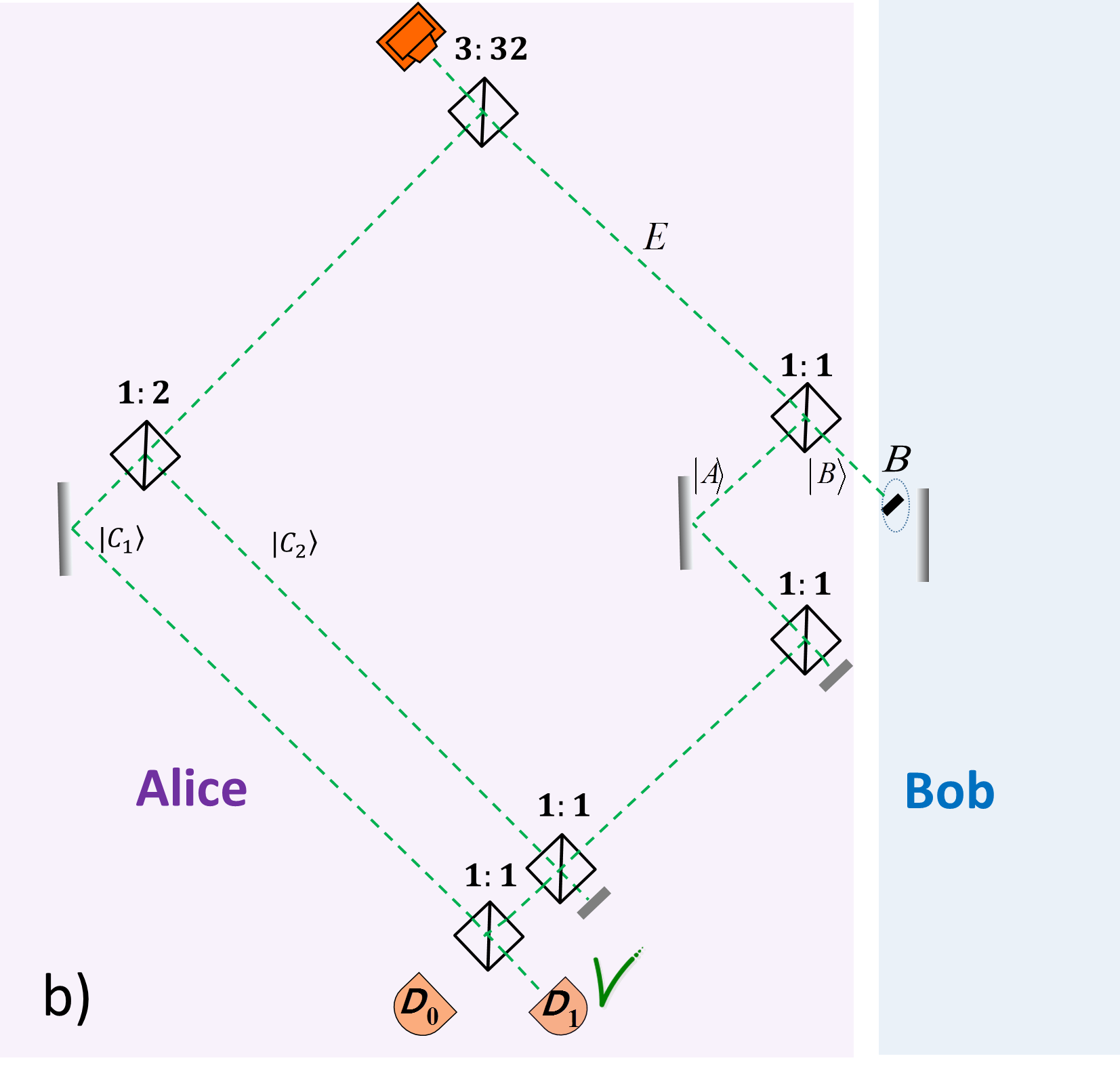}
  \caption{{\bf A counterfactual measurement protocol which is capable of verifying both presence and absence  of a block. }   a) When the small inner interferometer is empty, the probe leaves the right arm of the external interferometer. In this case, the second inner interferometer destructively interferes toward detector $D_1$. b) When $B$ is blocked, the port of detector $D_0$ is dark due to destructive interference. }
 \end{figure}

Similar failure of counterfactuality according to the weak trace appears in the  communication protocol described in Fig.~6,  which is counterfactual according the classical argument. There are two cases: the path $B$ is blocked and the path $B$ is free. When  the path is blocked, the probe does not leave any trace in $B$.

When the path is not blocked, we should consider two legitimate situations:  $D_0$ clicks, corresponding to the correct bit transmission and $D_1$ clicks, corresponding to the rare case of the erroneous bit transmission. Let us start with the the probable case of the click of $D_0$.

In this event, at the intermediate time, the photon in the interferometer is described by the forward-evolving state
\begin{equation} \label{psicomm}
\begin{split}
|\psi \rangle= \frac{1}{\sqrt {35}}\left(4|B\rangle+4|A\rangle+|C_{1}\rangle+\sqrt{2}|C_{2}\rangle\right),
\end{split}
\end{equation}
and the backward-evolving state
\begin{equation} \label{phicomm}
\begin{split}
\langle\phi|= \frac{1}{\sqrt{8}}\langle B|-\frac{1}{\sqrt{8}}\langle A|+\frac{1}{\sqrt{2}}\langle C_{1}|+\frac{1}{2}\langle C_{2}|.
\end{split}
\end{equation}
The corresponding weak value is $\left({\rm \bf P}_B\right)_w=1$, so the probability to find an orthogonal component is $\epsilon^2$.

When the path is free, but detector $D_1$ clicks in  error, the backward-evolving state is
\begin{equation} \label{ABLrule}
\begin{split}
\langle\phi|= -\frac{1}{\sqrt{8}}\langle B|+\frac{1}{\sqrt{8}}\langle A|+\frac{1}{\sqrt{2}}\langle C_{1}|-\frac{1}{2}\langle C_{2}|.
\end{split}
\end{equation}
This state is orthogonal to the preselected state $|\psi \rangle$, Eq. (\ref{psicomm}), so the weak value of ${\rm \bf P}_B$ becomes singular and cannot help with calculating the probability of finding the orthogonal state in the environment, given the $D_1$ error click. This probability depends on the details of the imperfections which led to the error. 

Let us assume that the reason for the error is the coupling with the environment in $B$ modeled by Eq. (\ref{eq::iaSingle}), and analyze both cases together. The (approximately normalized) state of the probe and the environment, postselected on the space of the probe passing the beamsplitter towards detectors $D_0$  and  $D_1$ is
\begin{equation}\label{simpleCFC}
 |0\rangle (\sqrt{1- \epsilon^2} |\chi \rangle + \epsilon |\chi^\perp \rangle) +|1\rangle ( (1-\sqrt{1- \epsilon^2})|\chi \rangle - \epsilon |\chi^\perp \rangle)  , 
\end{equation}
From this equation we get the two contributions to  the footprints of the probe in $B$. One contribution corresponds to a click at $D_0$ with an approximate probability 1 followed by finding $|\chi^\perp \rangle$, with probability $\epsilon^2$. The second contribution corresponds to a click at $D_1$ with probability $\epsilon^2$, followed by finding $|\chi^\perp \rangle$ with probability close to 1. In conclusion, the probability to detect an orthogonal component at the two legitimate detectors is $2\epsilon^2$.

The counterfactual communication protocol assisted by the quantum Zeno effect \cite{Ho06,Salih} consists of a chain of $M$ large interferometers, in every one of which a chain of $N$ small interferometers is inserted, see Fig.~4. The protocol works properly when $N\gg M\gg 1$. 

The trace in Bob's site is left only in the case of transmission of bit 0, when there is no block. To estimate this trace  we have to take into account both the click of $D_0$, happening with probability close to 1, and the click of $D_1$ (corresponding to an error) happening with a small probability, but resulting in a larger trace in Bob's site.

In this protocol we get $NM$ channels between Alice and Bob. In the channel $(m,n)$ (the $n$th small interferometer of the chain in the $m$th large interferometer) the forward-evolving state   is
\begin{flalign} 
|\psi \rangle &=\cos^{m}\frac{\pi}{2M}|C\rangle&&\\\nonumber
&+\cos^{m-1}\frac{\pi}{2M}\sin\frac{\pi}{2M}
 \left(\cos\frac{n\pi}{2N}|A\rangle+\sin\frac{n\pi}{2N}|B\rangle\right),&&
 \label{forward Salih}
\end{flalign}
where $|C\rangle=|C_m\rangle$ the state of the particle in the left arm of the $m$th large interferometer,  $|B\rangle=|B_{m,n}\rangle$ the state of the particle in the right arm of the $n$th  interferometer in the $m$th chain, etc.
 The backward-evolving state from $D_0$ is
 \begin{flalign}
\langle\phi|&= \cos^{M-m}\frac{\pi}{2M}\langle C|&&\\\nonumber
 &-\cos^{M-m-1}\frac{\pi}{2M}\sin\frac{\pi}{2M}
 \left(\sin\frac{n\pi}{2N}\langle A|-\cos\frac{n\pi}{2N}\langle B|\right).&& 
\end{flalign}
Thus, the weak value of projection on path $(m,n)$ is
 \begin{equation} \label{wvmn}
\begin{split}
 \left({\rm \bf P}_{m,n}\right)_w\simeq\frac{\pi^2}{8M^2}\sin\frac{n\pi}{N}.
\end{split}
\end{equation}
The probability to find at least some orthogonal component in Bob's site, when $D_0$ clicks, is the sum of probabilities for all paths
\begin{equation} \label{tracesalih}
 \sum_{m,n}|\left({\rm \bf P}_{m,n}\right)_w \epsilon~|^2\simeq\frac{\epsilon^2\pi^{4}N}{2^{7}M^3}.
\end{equation}

The backward-evolving wave function originating at  $D_1$ 
is significant at Bob's site only in the channels $(M,n)$ of the last large interferometer $M$: 
\begin{align}
\langle\phi| = \sin\frac{\pi}{2M}\langle C| 
  +\cos\frac{\pi}{2M}\left(\sin\frac{n\pi}{2N}\langle A|-\cos\frac{n\pi}{2N}\langle B|\right).
\end{align}
Thus, the weak value of projection on path $(M,n)$ in case of detection at $D_1$ is
  \begin{equation} \label{wvmn1}
\begin{split}
 \left({\rm \bf P}_{M,n}\right)_w\simeq-\frac{1}{2}\sin\frac{n\pi}{N},
\end{split}
\end{equation}
and summing on all paths in the chain $M$ given that $D_1$ clicks,
yields the probability to find at least some orthogonal component in Bob's site
\begin{equation}
 \sum_{n}|\left({\rm \bf P}_{M,n}\right)_w \epsilon~|^2\simeq\frac{\epsilon^2 N}{8}.
\label{trace prob D_1}
\end{equation}
Taking into account the probability of the click at $D_1$, $\frac{\pi^2}{4M^2}$, the total probability of finding an orthogonal component on Bob's site after a single legitimate run of the protocol is
\begin{equation} 
 {\rm Prob}=\frac{\epsilon^2 \pi^2 N}{2^5 M^2}\left(1+\frac{\pi^2}{4M}\right).
 \label{probtotal}
\end{equation}
For constant (large) ratio of $\frac{N}{M}$ but for very large $M$, the probability to find any orthogonal component is much smaller than $\epsilon^2$. So it seems that the protocol is counterfactual.
 However, in \cite{count} it was shown that the non-counterfactuality criterion we used, $|({\rm \bf P}_B)_w |$ of order 1, is not appropriate for protocols with multiple paths.

 \begin{figure}
  \includegraphics[width=0.95\linewidth]{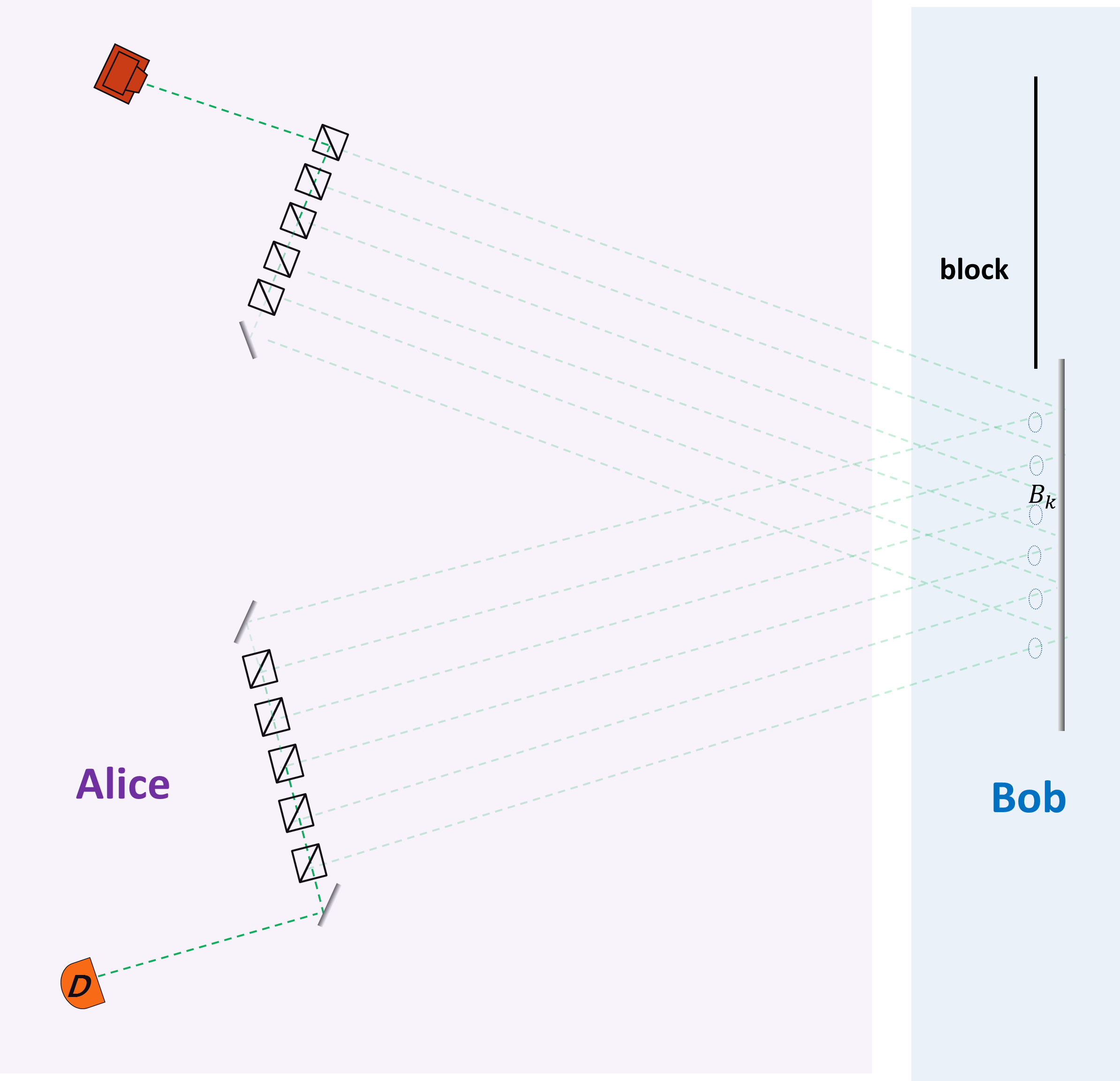}
  \caption{{\bf Non-counterfactual simple  communication protocol with $K$ paths. }   The wave function is split into a superposition of $K$ equal parts which bounce off Bob's mirror, who can block all of them with a large block. }
 \end{figure}

 Consider the following,  explicitly non-counterfactual, protocol with $K$ paths, Fig.~6, based on the single path protocol of  Fig.~1.  Bob has a single  mirror at his site which he can block for bit 1 and leave open for bit 0. An array of specially designed beam splitters divide the probe wave function to $K$ equal parts and another array  brings the reflected wave packets to a single wave packet moving towards Alice's detector. 
 The protocol is clearly not counterfactual when Alice detects the probe, since there is no way for the probe to reach Alice without visiting Bob's mirror. In our model all elements are ideal, but the parts of Bob's mirror become slightly entangled with the probe due to the local interactions of the probe at every path in the point where it touches Bob's mirror.  The local environment of every path $k$ at the mirror obtains a small component of the orthogonal state $|\chi^\perp\rangle_k$. After bouncing off the mirror, the quantum state of the probe and the environment is
\begin{equation} \label{Kpaths}
 \frac{1}{\sqrt K}\sum_{k=1}^K\prod_{j\neq k} |\chi\rangle_j~ (\sqrt{1-\epsilon^2}~|\chi\rangle_k +\epsilon |\chi^\perp\rangle_k)|k\rangle .
\end{equation}
At this stage, the  probability to find an orthogonal component, as in a single-path non-counterfactual protocol, equals $\epsilon^2$. But we consider the postselected case, when Alice gets the click. The postselection is on the spatial state $ \frac{1}{\sqrt K}\sum_{k=1}^K |k\rangle $, the probability for which is close to 1 for large $K$ and small $\epsilon$. However, after the postselection, the probability to find the orthogonal component reduces by a factor of $\frac{1}{K}$. 
In the protocol we discuss, the number of paths is $K=MN$, so the probability Eq. (\ref{probtotal}) is much larger than   $\frac{\epsilon^2}{K}$, the probability to find an orthogonal component in  a protocol with a photon actually traveling between Alice and Bob. Therefore, we cannot claim that the protocol is counterfactual.

 The last protocol we want to analyze here using the trace criterion is 
 a simple Zeno-type communication protocol suggested by A-SB which is just one inner chain with $N$ interferometers of the  protocol described in Fig.~4, but with some modified rules. Following A-SB we  consider a particular geometry of Alice's territory, see Fig.~7.
 As usual, blocking all channels leads to a click of Alice's detector which tells Alice that the bit is 1 without leaving a trace on Bob's mirror. 
 Obtaining no click tells us that the bit is 0. Of course, in this case there is a  large trace in Bob's site, the probability to find an orthogonal component on Bob's mirrors is
 \begin{equation} \label{ABLrule}
 \sum_{n=1}^{N-1}\epsilon^2\sin^{4}\frac{n\pi}{2N}\simeq\frac{3N\epsilon^2}{8},
\end{equation}
and the presence of Alice's photon in Bob's site clearly makes the quantum state of Bob's site orthogonal to its original state.
  
 \begin{figure}
  \includegraphics[width=\linewidth]{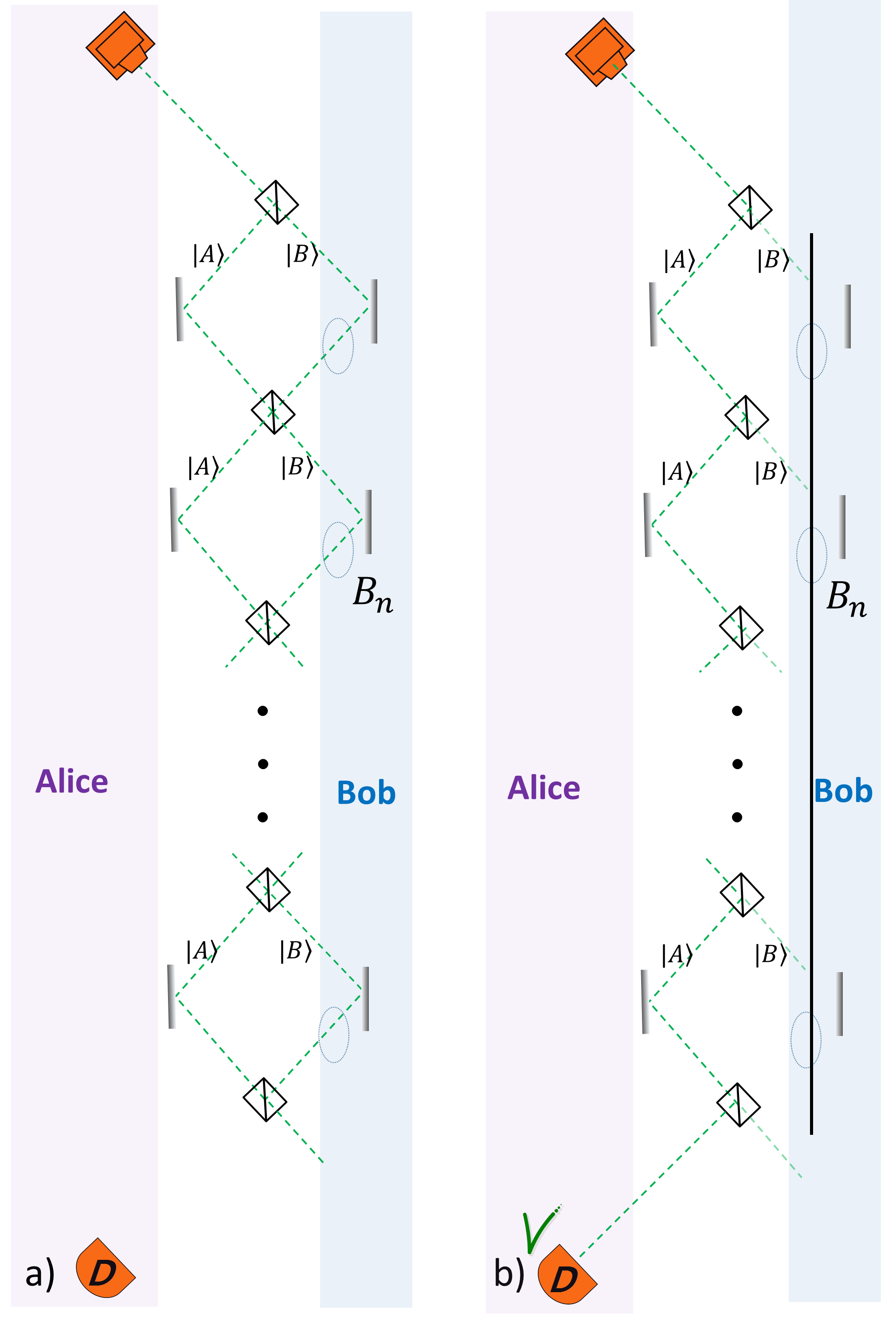}
  \caption{{\bf The A-SB protocol.} a) The Zeno chain of $N$ identical MZIs brings the probe  outside the interferometer when it is ideal and empty. 2) When Bob blocks all  his mirrors, the probe  reaches Alice's detector with probability close to 1 up to a term of order $\frac{1}{N}$ . }  
   \end{figure}

 A-SB argue that since in this case the photon moves from Alice to Bob, while the information is transferred from Bob to Alice, the trace in Bob's site should be disregarded. We do not support this approach, but there is no need to argue about it because 
 when the paths are not blocked, there is also a significant probability to find an orthogonal component of the environment in Bob's site when Alice's detector does click (in error transmission). The probability of the error is  small
 \begin{equation} \label{ABLrule}
\begin{split}
 \sum_{n=1}^{N-1}\frac{\epsilon^2}{4}\sin^2\frac{n\pi}{N}
 \simeq\frac{N\epsilon^2}{8},
 \end{split}
\end{equation}
but we do get a significant contribution to the trace in Bob's site, because, when Alice gets the click, the probability to find some orthogonal component in Bob's site is approximately 1, so the probability of the error click is also the probability of finding the trace in Bob's site. 
Thus, running the  protocol for bit 0, leads to the probability of finding an orthogonal component in Bob's site after the end of the procedure,   $\frac{N\epsilon^2}{2}$. This is clearly not counterfactual. (For a protocol with $N$ channels the probability has to be much smaller than $\frac{\epsilon^2}{N}$). 

An interesting proposal which involves both spatial and polarization degrees of freedom of the probe \cite{SalRar}  has a similar weakness. It can be argued that it is completely counterfactual when transmission is correct, but it has a small probability for error clicks, which in addition to sending wrong bits are sent in a non-counterfactual manner. However, since these clicks are legitimate events according to the protocol, they cannot be dismissed.

For completeness of the weak trace analysis, we note an apparently relevant work titled ``Quantum counterfactual communication without a weak trace'' \cite{Arvid}. In spite of the title, the described protocol does leave a weak trace according  to the definition of the current paper. The reason for the title is an extreme weakness of the coupling in the experiment, but it makes the reference non-counterfactual trace very weak too, so  according to the criterion of the weak trace which considers the ratio, the procedure does have a weak trace.

\section{The Fisher information criterion}

A-SGB proposed a somewhat different approach to analyze counterfactuality.  They suggested quantifying the presence of particles in Bob's site by the information about some properties of Bob's site transferred to the particles during the protocol. It may seem that there is a conceptual contradiction within this approach. The counterfactual communication protocol is supposed to bring information about Bob's site (presence of Bob's block in a particular place) to Alice without particles being there. So, if Alice gets the information, then according to the A-SGB definition, the particle was there. However, the  information to be communicated between Bob and Alice  can be  separated from the information about other aspects of Bob's site in an unambiguous way, so the A-SGB approach might shed light on the counterfactuality of various protocols.

As we discussed in Section II, we do not accept A-SB's general strategy of evaluating counterfactuality by calculating Fisher information which is obtained from all possible outcomes of an experiment, i.e., calculation of Fisher information without postselection. We have argued that the existence of a counterfactual communication protocol without postselection is impossible. It is the act of discarding some of the cases that makes quantum counterfactual communication possible. Quantum counterfactual communication with postselection achieves significantly more than classical counterfactual communication with prior agreement presented by A-SB. Moreover, the general result obtained by A-SGB strengthens our claim. They showed that the analysis of Fisher information without postselection is equivalent to the analysis based on the density of the forward-evolving wave function. It would be against any physical intuition if we could get information about a particular location when the wave function of our probe was not present there.
However, we do adopt A-SGB's idea of evaluating counterfactuality through analysis of Fisher information about variables in Bob's site which is  present in the probe particles  reaching Alice's detectors. 

Let us define more precisely the rules of our analysis.
To simplify the calculations, we will make the following (not necessarily realistic) assumptions. All optical devices  are ideal except for the distortion of polarization at Bob's site given by rotation in a fixed basis characterized by the parameter $\theta$. The Fisher information about $\theta$ will then be considered for assessing the counterfactuality of each protocol. We assume that  the initial polarization state  
is always $|H\rangle$ and the distortion is described by the following transformation of  the states 
\begin{align}\label{distortion}
|H\rangle \rightarrow \cos\theta|H\rangle+ \sin\theta|V\rangle,~~ \nonumber\\
 |V\rangle \rightarrow -\sin\theta|H\rangle+\cos\theta|V\rangle.
\end{align}
The estimate of counterfactuality will be the ratio of the Fisher information obtained in the protocol and the Fisher information obtained in the reference non-counterfactual procedure described in Fig.~1, when Bob does not put a block in $B$, but the polarization of the photon is rotated in $B$ according to Eq. (\ref{distortion}).
The Fisher information about $\theta$ might be a complicated function of $\theta$. For a discrete random variable with outcomes $i$, given a conditional probability  $P(i|\theta)$, it is
\begin{equation} \label{Fisher}
\begin{split}
F_X(\theta)=\sum\limits_{i}^{}P(i|\theta)\left[\partial_{\theta}
(\log P(i|\theta))\right]^2 
=\sum\limits_{i}^{}\frac{\left[\partial_{\theta}(P(i|\theta))\right]^2}{P(i|\theta)}.
\end{split}
\end{equation}
As it is usually done, we  consider the ratio in the limit of small distortion, i.e.  at  the limit  $\theta \rightarrow 0$.
At the reference scenario the Fisher information is
\begin{equation}\label{Fisherref}
F_{\rm ref}=\lim\limits_{\theta \to 0}\left[\frac{\left(\partial_{\theta}P(V|\theta)\right)^2}{P(V|\theta)}+\frac{\left(\partial_{\theta}P(H|\theta)\right)^2}{P(H|\theta)}\right]=4.
\end{equation}

The EV IFM of finding  the block is  fully counterfactual, since the  detected probes are not distorted by the rotation of polarization at $B$. The state of the detected particles is simply $|H\rangle$, and therefore,  trivially, the obtained Fisher information is zero.

In the IFM  protocol for the absence of the object Fig.~3, for our model with distortion only at $B$, the polarization state of the probe at detector $D$ is  exactly the same as  if the photon were passing only through $B$ because the wave packets passing through paths $C$ and $A$ interfere destructively towards $D$. 
Thus, the Fisher information is exactly the same as  in the case of the reference probe,   $F=4$. The protocol is not counterfactual.

In the protocol described in Fig.~6, when the transmitted bit is 1, i.e. there is a block in $B$, there is no distortion of the polarization of the photons reaching Alice's detectors. Therefore, for bit 1 the Fisher information is zero and the protocol is counterfactual. But of course, the difficulty arises for bit 0, when $B$ is not blocked. In this case,  the (approximately normalized for small $\theta$) state of the probe, including its polarization, postselected on the subspace of the probe wave packets passing through the final beamsplitter  towards detectors $D_0$  and  $D_1$ is
\begin{equation}\label{eq::iaSingletheta}
 |0\rangle (\cos \theta  |H \rangle + \sin \theta |V \rangle) +|1\rangle ( (1-\cos \theta)|H \rangle -  \sin \theta  |V \rangle)  , \end{equation}
Now we have four discrete outcomes for the Fisher information analysis: {$i=$ $0H$,  $0V$,  $1H$, and $1V$} corresponding to the states $ |0\rangle  |H \rangle$, $ |0\rangle  |V \rangle$, etc.  The straightforward calculation yields
\begin{equation} \label{Fisher1}
\begin{split}
F
=\lim\limits_{\theta \to 0}
\sum\limits_{i}^{}\frac{\left[\partial_{\theta}(P(i|\theta))\right]^2}{P(i|\theta)}=8.
\end{split}
\end{equation}
Thus, also according to the Fisher information criterion, the protocol is not counterfactual.

For the analysis of the protocols which include the quantum Zeno effect we have to specify how to deal with the fact that the probe passes several regions in Bob's site. We will follow A-SB
and assume that the distortion $\theta$ happens only in one path, but then we sum the Fisher information of all paths. 
Later we will consider another approach according to which the distortion $\theta$ is  the same for all paths which is relevant for some realistic implementations.

In counterfactual communication protocols employing the Zeno effect \cite{Ho06,Salih} without blocks, the Fisher information obtained about the distortion at the $n$th small interferometer of the chain in the $m$th large interferometer 
can be estimated in the following way. For $N\gg M\gg 1$, required in this protocol, the  amplitude of the $|H\rangle$ polarization state at $D_0$ is approximately
\begin{equation} \label{m,n+H}
1- \frac{\pi^2}{8M} - \theta^2~\frac{\pi^2}{8M^2} \sin \frac{n\pi}{2N}~\sin \frac{(N-n)\pi}{2N}.
  \end{equation}
The amplitude of polarization $|V\rangle$  at $D_0$ is approximately  
\begin{equation} \label{m,n+V}
 \theta~\frac{\pi^2}{4M^2} \sin \frac{n\pi}{2N}~\sin \frac{(N-n)\pi}{2N}.
  \end{equation}
Using 
 Eq. (\ref{Fisher})  we  calculate the  contribution to Fisher information of the probe reaching detector $D_0$ due to distortion in channel $(m,n)$  
\begin{equation} \label{m,n}
 F_{m,n}=\frac{\pi^4}{2^4 M^4}\sin^{2}\frac{n\pi}{N} .
\end{equation}
The summation of the information obtained for all paths yields  
\begin{equation} \label{summ,n}
 F=\frac{\pi^{4}N}{2^{5}M^3}. 
\end{equation}

Significant Fisher information of the probes reaching  $D_1$ (corresponding to erroneous clicks) comes only from the distortions at the last large interferometer $M$. The  amplitude of polarization $|H\rangle$ at $D_1$, due to distortion in channel $(M,n)$, is 
\begin{equation}
\frac{\pi}{2M}-\frac{\pi^3}{8 M^2} - \theta^2~\frac{\pi}{4M} \sin \frac{n\pi}{2N}~\sin \frac{(N-n)\pi}{2N}.
\end{equation}
The amplitude of polarization $|V\rangle$  at $D_1$ is 
\begin{equation}
 -\theta~\frac{\pi}{2M} \sin \frac{n\pi}{2N}~\sin \frac{(N-n)\pi}{2N}.
\end{equation}
Thus, the  contribution to Fisher information due to distortion in channel $(M,n)$ from detector $D_1$  is
\begin{equation}
 F_{M,n}=\frac{\pi^2}{4 M^2}\sin^{2}\frac{n\pi}{N} .
\end{equation}
The summation on  all paths of the last external MZI yields
\begin{equation}
 F=\frac{\pi^{2} N}{8 M^2}. 
 \label{Fisher Salih D_1}
\end{equation}
The total Fisher information obtained from the two detectors 
is 
\begin{equation}
 F=\frac{\pi^{2} N}{8 M^2}\left(1+\frac{\pi^2}{4M}\right). 
 \label{Fisher Salih total}
\end{equation}

If the ratio  $\frac{N}{M} \gg 1 $ but fixed, and we make $M$ very large, the Fisher information goes to zero.
However, we can claim (analogously to the claim for the weak trace) that the Fisher information criterion has to be modified when the channel has multiple paths. 

Consider the non-counterfactual protocol described in Fig.~6 with $K$ paths. Now we assume that the probe initially has a polarization state $|H\rangle$ and when it touches Bob's mirror, it transforms according to Eq. (\ref{distortion}). Again, we  follow the A-SB approach according to which we assume that it happens only in one path and  then multiply by the number of paths. If the distortion Eq. (\ref{distortion}) happens, say, in path $j$, then the probe, given that it reaches Alice's detector, will have the polarization state 
\begin{equation}\label{justj}
\sqrt{1-\frac{\sin^2 \theta}{K^2}}   |H \rangle + \frac{\sin \theta}{K} |V \rangle.
 \end{equation}
The Fisher information from this path, applying Eq. (\ref{Fisher}), is $F_j=\frac{4}{K^2}$. After summation on all paths we get $F=\frac{4}{K}$. Thus, also  according to the Fisher information approach, the reference for a protocol with $K$ path obtains a factor of $\frac{1}{K}$. In the Zeno type protocol we discuss,  the number of channels is $K=MN$, so 
\begin{equation}\label{justj}
F\approx\frac{\pi^{2} N^2}{8 M}\frac{1}{K}\gg \frac{4}{K},
 \end{equation}
which shows that the ``counterfactual communication'' protocols in \cite{Ho06,Salih} are not counterfactual. 

The last setup we consider is the A-SB  protocol described in Fig.~7. 
The information obtained at Bob's site, when he does not put the blocks and Alice does not get the click, is
\begin{equation} \label{ABLrule}
F=  \sum_{n=1}^{N-1}4\sin^{4}\frac{n\pi}{2N}\approx\frac{3N}{2},
 \end{equation}
but we find it irrelevant since the probe does not reach Alice.

The overall Fisher information obtained about all the paths of Bob at Alice's detector is 
\begin{equation} \label{ABLrule}
\begin{split}
F= \sum_{n=1}^{N-1}\sin^2\frac{n\pi}{N}
 \approx\frac{N}{2}.
 \end{split}
\end{equation}
This is  much bigger than the reference $F=4$ and, of course, much bigger than the reference $\frac{4}{N}$, which should be taken when we have $N$ paths in the channel between Alice and Bob. Thus, the protocol is clearly not counterfactual according to our definition.

\section{Coherent footprints}

For a protocol with multiple paths, it is of interest to consider another model of interaction. An actual implementation of counterfactual communication using  the quantum Zeno effect \cite{Ho06,Salih} used a simplified geometry in which the multiple mirrors on Bob's side were implemented by just one mirror with the probe bouncing off  at different times, see Fig.~8. In the model we considered before, at every path there was a separate system on which the particle left the trace, see Fig.~4.

\begin{figure}
  \includegraphics[width=0.95\linewidth]{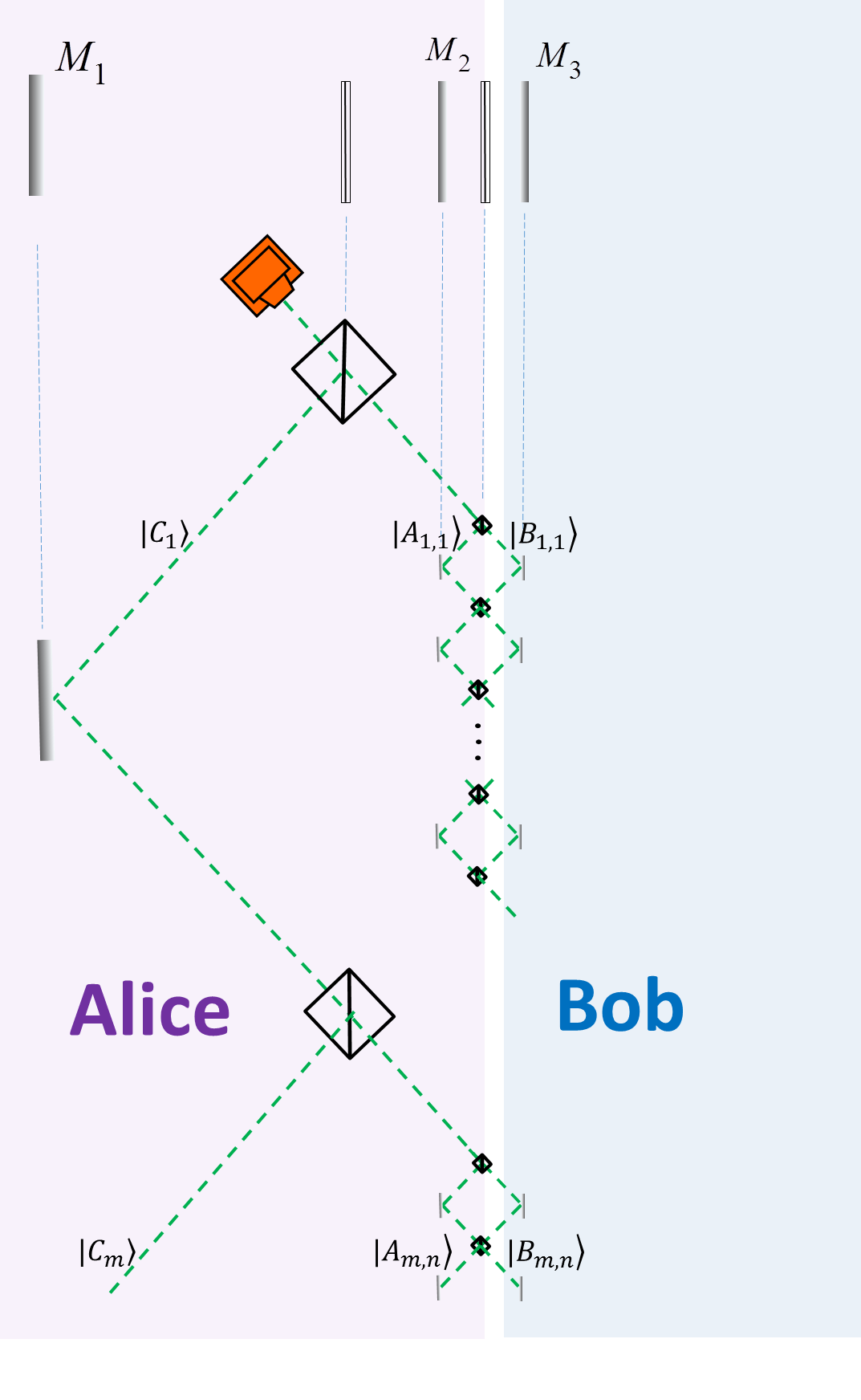}
  \caption{
  {\bf Coherent Zeno-type counterfactual communication \cite{Ho06,Salih}. }   For a horizontally moving probe three mirrors and two beamsplitters can replace the multiple mirrors and beamsplitters of the interferometer described in  Fig.~4 (partly reproduced here). A gedanken observer moving  upward  with the velocity of the probe sees the experiment of Fig.~4. The modification requires a mechanism (not shown) which removes  $M$ times the mirror $M_2$ to allow the probe entering and leaving the inner interferometer after bouncing $N$ times inside. It also requires a special mechanism  (not shown) of putting the probe in, and another mechanism (not shown) measuring it at the end by inserting the detectors. Mirror $M_3$ accumulates coherently the contributions of interaction with the probe at different times. 
  }
  \label{fig:setup}
\end{figure}

A simple protocol which demonstrates the difference is the  situation when the probe bounces $K$ times from Bob's mirror, see Fig.~9. When the local environment at every  meeting of the path with Bob's mirror creates an orthogonal component, the process causes the following change of the quantum state of the environment:
\begin{equation} \label{Kpaths1}
 \prod_{k=1}^K |\chi\rangle_k \rightarrow \prod_{k=1}^K (\sqrt{1-\epsilon^2}~|\chi\rangle_k +\epsilon |\chi^\perp\rangle_k) .
\end{equation}
The probability to find an orthogonal state at Bob's site is $K\epsilon^2$.

\begin{figure}
  \includegraphics[width=0.95\linewidth]{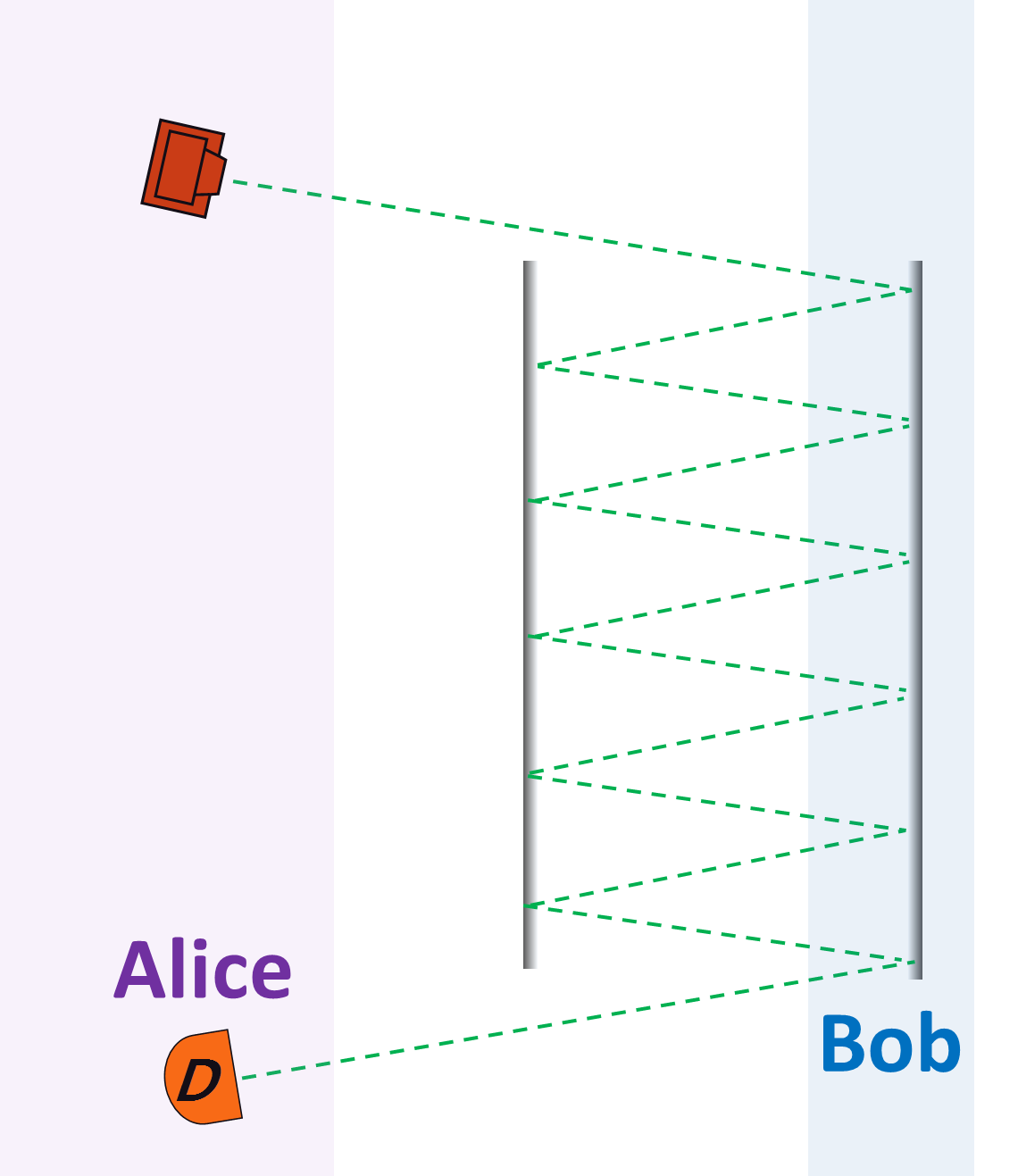}
  \caption{
  {\bf Coherent and incoherent  footprints of the probe.} A non-counterfactual communication protocol with repeating passage of the probe. The footprints of every passage can be considered incoherent when they are local disturbances at different locations of Bob's mirror, or coherent, when they provide the change of the total momentum of Bob's mirror.
  }
  \label{fig:setup}
\end{figure}

The obvious way that the probe leaves a trace on the environment is the momentum  transfer to a mirror when it bounces off. Then, if it bounces $K$ times, the momentum which the probe  transfers is approximately $K$ times bigger.  The mirrors in the interferometer have to be well localized and we can approximate the quantum state of a mirror in the momentum representation as a Gaussian. Then, if one bounce leads to Eq. (\ref{eq::iaSingle}), then $K$ bounces will lead (for small $\epsilon$) to
\begin{equation}\label{eq::iaSinglew}
|\chi \rangle \rightarrow \sqrt{1-K^2 \epsilon^2} ~|  \chi \rangle +  K \epsilon ~ |\chi^\perp \rangle .
\end{equation}
The probability to find an orthogonal state at Bob's site is therefore $K^2\epsilon^2$, it is increased by a factor of $K$ in the case of coherent footprints in comparison to the case of incoherent footprints Eq. (\ref{Kpaths1}).

The reference case of the non-counterfactual protocol with $K$ paths, Fig.~6, is changed too. Again, if the model of interactions is the transfer of the momentum of the probe  to the mirror, then every part $|k\rangle$ of the probe transfers the same momentum to the mirror. The probe in the superposition of being in different paths transfers the momentum as if it was just in one path. Thus, instead of Eq. (\ref{Kpaths}), the quantum state of the probe and the environment, after the probe wave packets bounced off the mirror, is 
\begin{equation} \label{Kpaths2}
 \frac{1}{\sqrt K}(\sqrt{1-\epsilon^2}~|\chi\rangle +\epsilon |\chi^\perp\rangle)\sum_{k=1}^K |k\rangle .
\end{equation}
There is no entanglement between the spatial wave function of the  probe and the environment, so the method of reducing the trace by choosing a postselected state with particularly high probability of success \cite{count} does not work. 

To calculate the  weak trace, i.e., the probability of finding an orthogonal component in the environment (Bob's single mirror) in the coherent model we can use the same expression Eq. (\ref{wvmn}) for $\left({\rm \bf P}_{m,n}\right)_w$, but  replace Eq. (\ref{tracesalih}) 
by
\begin{equation} \label{tracesalihcoherent}
| \sum_{m,n}\left({\rm \bf P}_{m,n}\right)_w \epsilon~|^2\simeq\frac{\epsilon^2\pi^{2}N^2}{2^{4}M^2},
\end{equation}
and use  Eq. (\ref{wvmn1}) for $\left({\rm \bf P}_{M,n}\right)_w$ (for $D_1$ click)   and replace Eq. (\ref{trace prob D_1}) by
\begin{equation} 
|\sum_{n}\left({\rm \bf P}_{(M,n)}\right)_w \epsilon~|^2\simeq\frac{\epsilon^2 N^2}{\pi^2}.
\end{equation}
Taking into account the probability of the detector clicks, we obtain the probability of finding the orthogonal component
\begin{equation} 
 {\rm Prob}=\frac{\epsilon^2 N^2}{4M^2}\left(1+\frac{\pi^2}{4}\right).
 \label{probtotalcoherent}
\end{equation}
Hence, even without the correction factor of multiple paths, which is not applicable here, the protocol is not counterfactual.

Coherent footprints are even more natural to consider with the Fisher information criterion for counterfactuality since the probe usually does not have many degrees of freedom. The standard approach of A-SB is to calculate the Fisher information obtained for one path and multiply by the number of paths. This corresponds to different distortions of the probe at different paths. If we have a system as in Fig.~8, it is natural to expect that the same rotation of polarization Eq. (\ref{distortion}) will happen at every path. 

The weak value formalism can help to perform the Fisher information calculation. 
For small $\theta$ ($ \ll \frac{M }{N}$) the angle of polarization rotation can be estimated as $\theta$ multiplied by the sum of weak values of projections of interaction regions. For   $D_0$ click, $\left({\rm \bf P}_{n,m}\right)_w$ are given by Eq. (\ref{wvmn}) so, the rotation angle is 
\begin{equation} 
\sum_{m,n}\left({\rm \bf P}_{m,n}\right)_w~\theta= \frac{\pi N\theta}{4M}.
\end{equation}
For $D_1$ click, only rotations in the last chain of the inner interferometers are relevant, so using Eq. (\ref{wvmn1}) we obtain
\begin{equation} 
\sum_{n}\left({\rm \bf P}_{M,n}\right)_w~\theta= \frac{ N\theta}{\pi}.
\end{equation}
Thus, the wave function passing through the final beamsplitter towards detectors $D_0$ and $D_1$ can be expressed approximately as 
\begin{flalign} \nonumber
|0\rangle\left(\cos\frac{\pi N \theta}{4M}~|H\rangle+\sin\frac{\pi N\theta}{4M}~|V\rangle\right) +~~~~~~~~~~ \\
|1\rangle~\frac{\pi}{2M}\left(\cos\frac{ N\theta}{\pi}~|H\rangle + \sin\frac{ N\theta}{\pi}~|V\rangle\right) .
\end{flalign}
Then we calculate, based on Eq. (\ref{Fisher}), the   Fisher information obtained from the two detectors at Alice's site:
\begin{equation}
F= \frac{N^2}{M^2}\left(1+\frac{\pi^2}{4}\right).
\end{equation}
This is much larger than the reference Fisher information, so the protocol is  not counterfactual.





\section{Aharonov-Vaidman modification of counterfactual protocols}

All the above apparently leads to a conclusion that apart from the classical criterion for counterfactuality, which gives rise to a contradiction and should be abandoned, there is no counterfactual protocol for testing that a particular place is empty. Thus, one may think that there is no protocol for counterfactual communication of a classical message, and therefore, the there is no counterfactual communication of a quantum state \cite{Li15,salih16,salih20,V16} which is based on it. The weak trace and Fisher information criteria showed that all the proposed protocols for counterfactual communication are actually not counterfactual as they leave a trace at Bob's site which is larger than the probe actually visiting Bob's site, or they allow for the acquisition of more information from the probe than the information which can be learned from a probe which has actually been at the site.

\begin{figure}
  \includegraphics[width=0.95\linewidth]{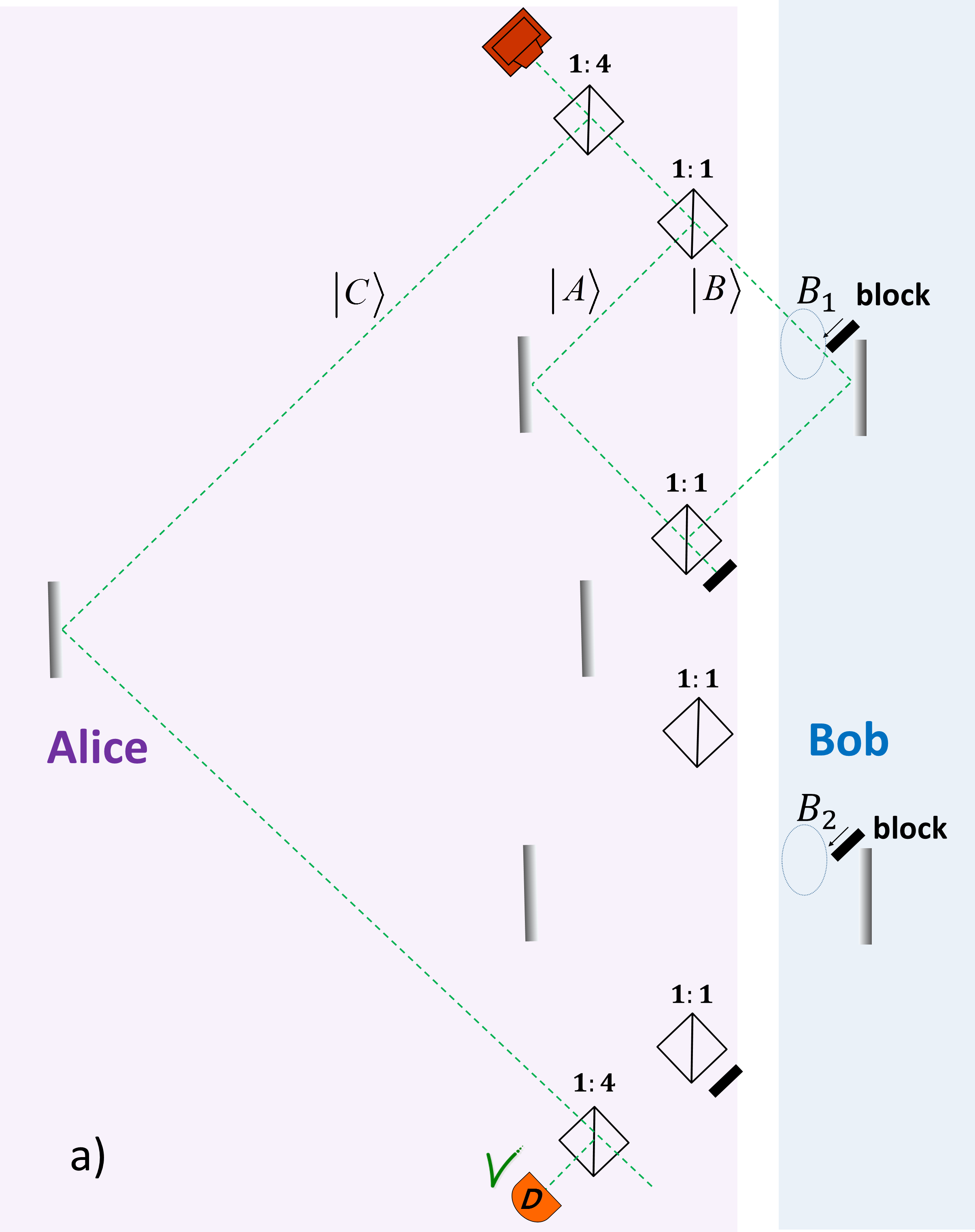}
  
   \includegraphics[width=0.95\linewidth]{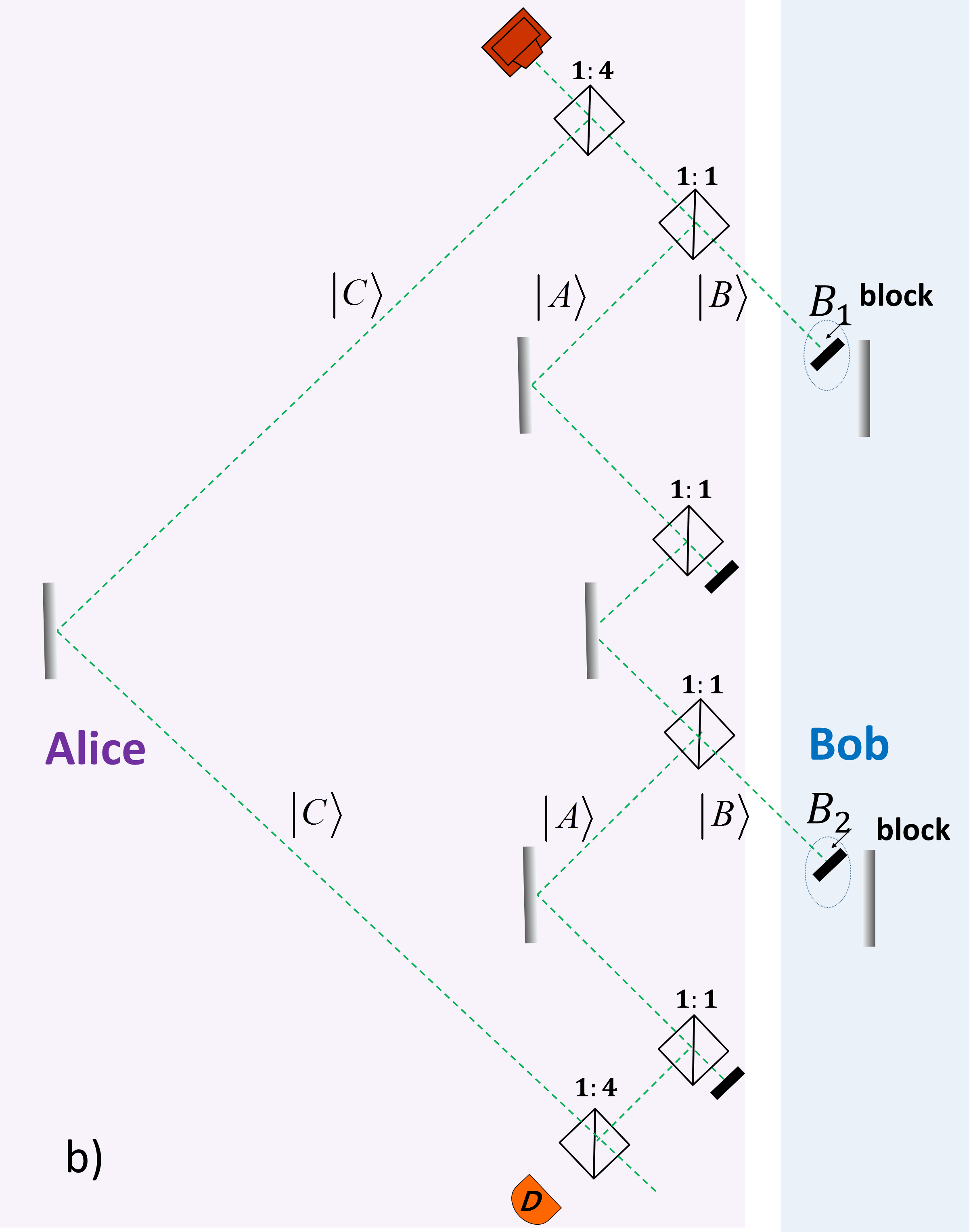}
  \caption{{\bf Counterfactual measurements of  the  absence  of two blocks without leaving a trace. }   a) Two consecutive small MZIs inserted in the right arm of the interferometer are tuned to send the probe out of the large interferometer, if they are empty.   b) When the right arms of the small interferometers are blocked by Bob, there is destructive interference at  detector $D$. }
 \end{figure}

The TSVF suggests that a counterfactual measurement showing a particular place to be empty  is impossible. Assume that it is possible. A probe provides the information that the place $B$ is empty and no weak trace is left in $B$. The way Alice gets the information is a click. It should not be possible if $B$ is blocked. This means that the forward-evolving wave function of the probe must be modified by the block, i.e. due to locality of interactions the wave must be in $B$. Moreover, the wave packet at $B$ must reach the point of the click at Alice's detector. If it does not, putting a block will not affect the probability of Alice's click. But if the wave packet from $B$ reaches the detector, the backward-evolving state from the detector reaches $B$. If we have an overlap of forward- and backward-evolving states, the weak value of the projection of the probe on $B$ is finite, so the weak trace is finite, and thus, the protocol is not counterfactual.

The idea of Aharonov and Vaidman (AV) \cite{AV19} is to consider two space-time points $B$ at Bob's site,  $B_1$ and  $B_2$, one after the other. In the communication protocol, Bob either blocks both or leaves both open. 
Now, we need that when the the two blocks  modify the forward-evolving wave function, then Alice's click becomes impossible. Therefore, the forward-evolving wave function must be at points $B$ in order to be influenced by the blocks. But it does not mean that the undisturbed forward-evolving wave function must be there: we need it to reach $B_1$, and when it is blocked, it also reaches $B_2$. It does not have to reach $B_2$ when $B_1$ is not blocked. To affect the probability of Alice's click, the wave packet from $B_2$ must reach the detector, but the backward-evolving state from the detector need not necessarily reach $B_1$. 

This idea is implemented in the scheme presented in Fig.~10. This is a MZI with two nested MZIs in the right arm. When the MZI is free, the forward-evolving state enters only the first inner MZI, and the backward-evolving state enters only the second. Bob's places $B_1$ and  $B_2$ are inside these interferometers, and since there is no overlap of the forward- and backward-evolving states inside inner MZIs in case of the click of $D$, we get no weak trace there, see Fig~ 10b. On the other hand, the interferometer is tuned in such a way that when Bob blocks $B_1$ and  $B_2$, $D$ cannot click, so its click unambiguously tells us that $B_1$ and  $B_2$ are empty.

Let us analyze the counterfactuality of the AV scheme, Fig.~10, according to the Fisher information criterion. If we apply the standard A-SB approach for the multiple paths case, i.e. assume polarization rotation only in one place and then sum the  contributions of all paths, we trivially find that the protocol is counterfactual. Polarization rotations spoil the balance of the inner interferometer, but if it is assumed that it happens only in one inner interferometer, then  no orthogonal polarization component $|V\rangle$ reaches Alice's detector. Thus, the Fisher information calculation will yield zero.

The AV idea can be applied to other counterfactual communication protocols to remove the weak trace of the probe from Bob's site. To modify the counterfactual communication described in Fig.~6, we just replace the MZI on the right arm of the large interferometer, by two sequential MZIs, see Fig.~11.

\begin{figure}
  \includegraphics[width=0.9\linewidth]{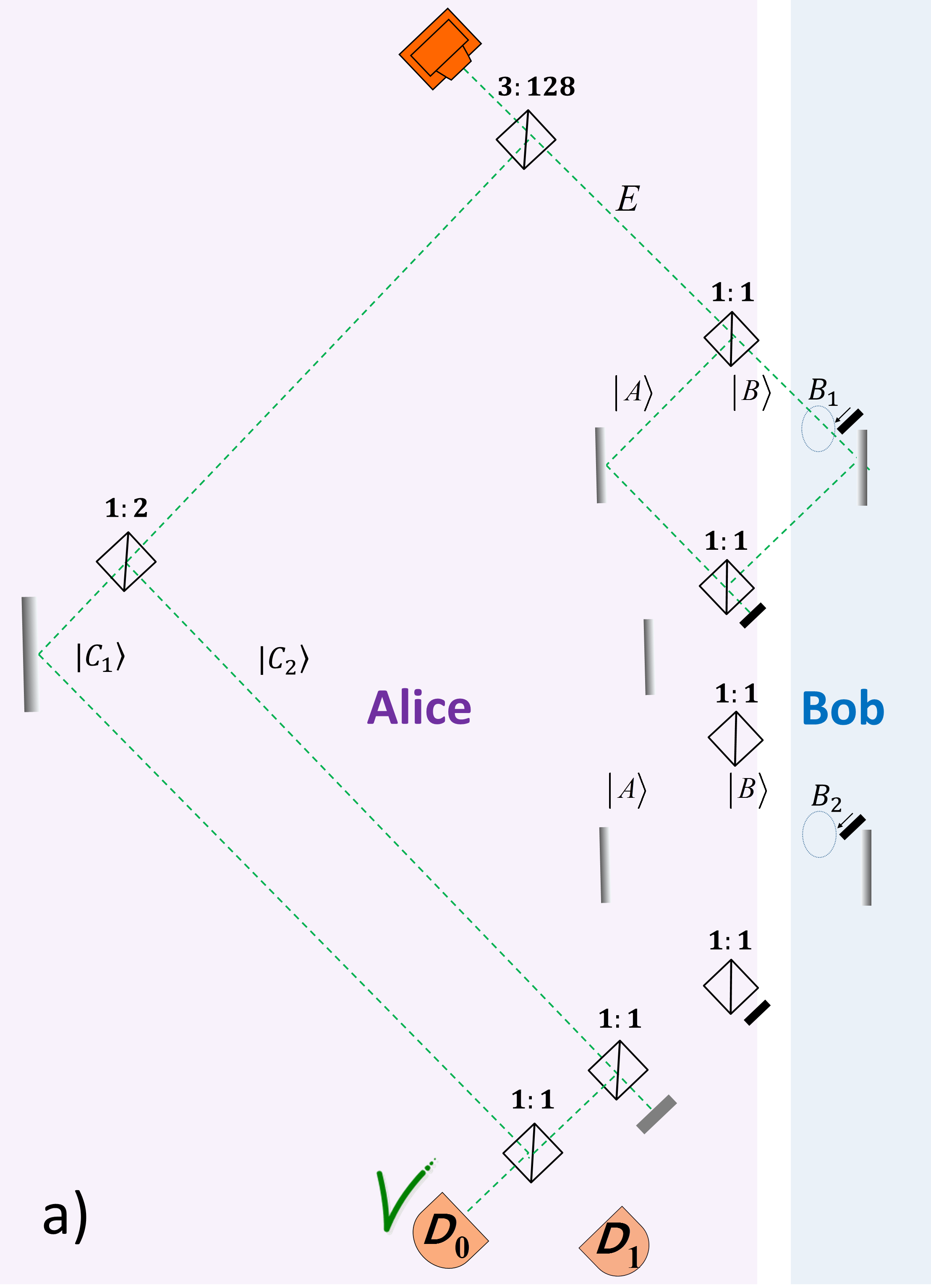} 
   \includegraphics[width=0.9\linewidth]{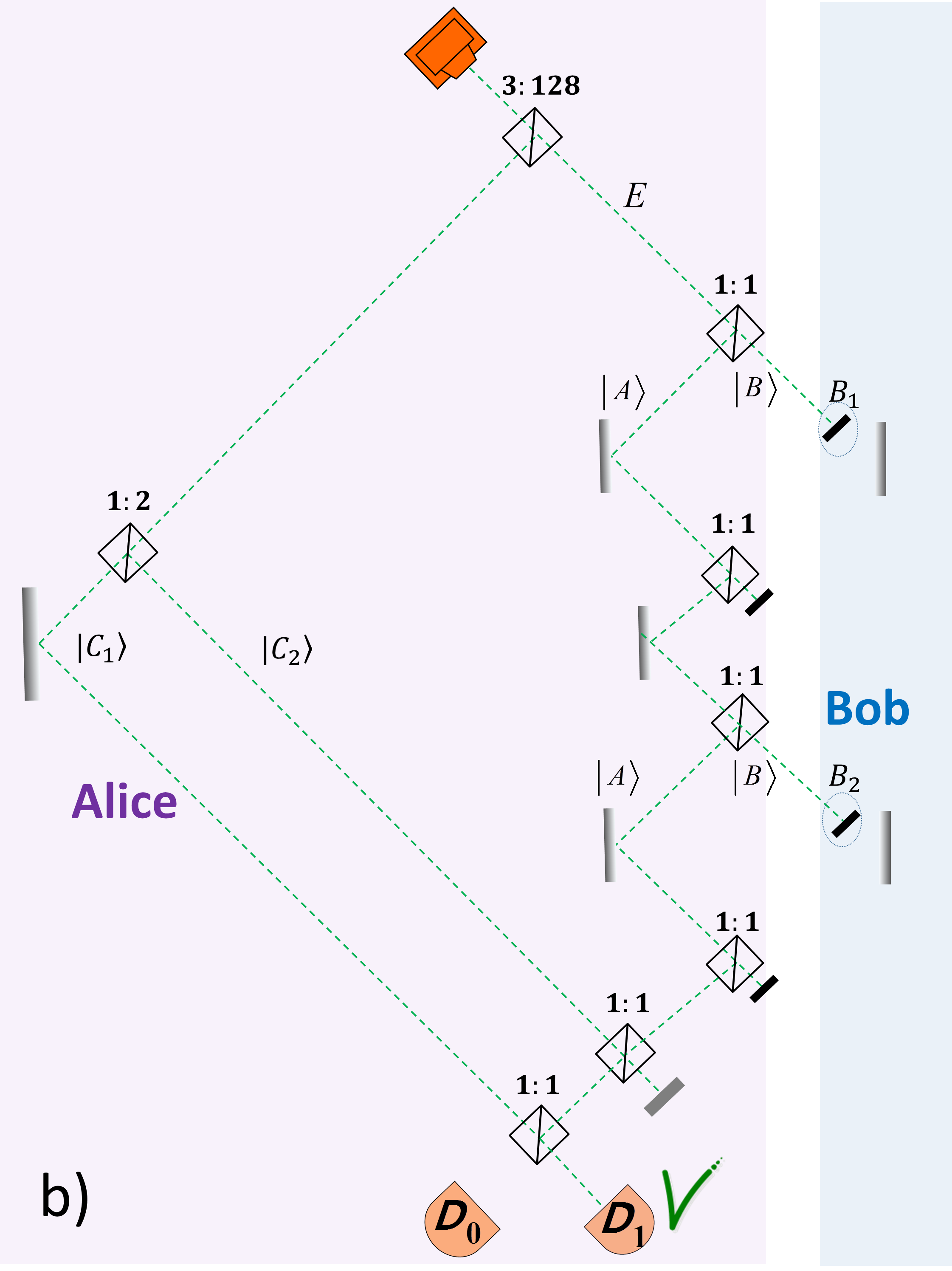}
  \caption{{\bf Modification of the protocol described in Fig.~4 which removes the weak trace.}
    a) Two consecutive small MZIs inserted in the right arm of the interferometer. Both are tuned to send the probe out of the large interferometer if they are empty. In this case there is destructive interference toward $D_1$.  b) When the right arms of the small interferometers are blocked by Bob, there is destructive interference toward $D_0$. }
 \end{figure}

To remove the trace in the protocol described in Fig.~4,  we need essentially to double every large MZI which includes the chain of inner MZIs, introducing two chains of inner interferometers, see Fig.~12.  Then, the forward-evolving wave will enter just one inner chain, while the backward-evolving wave will enter another. Without overlap on Bob's site, we do not get the first order trace as a single localized probe would get.

\begin{figure}
  \includegraphics[width=\linewidth]{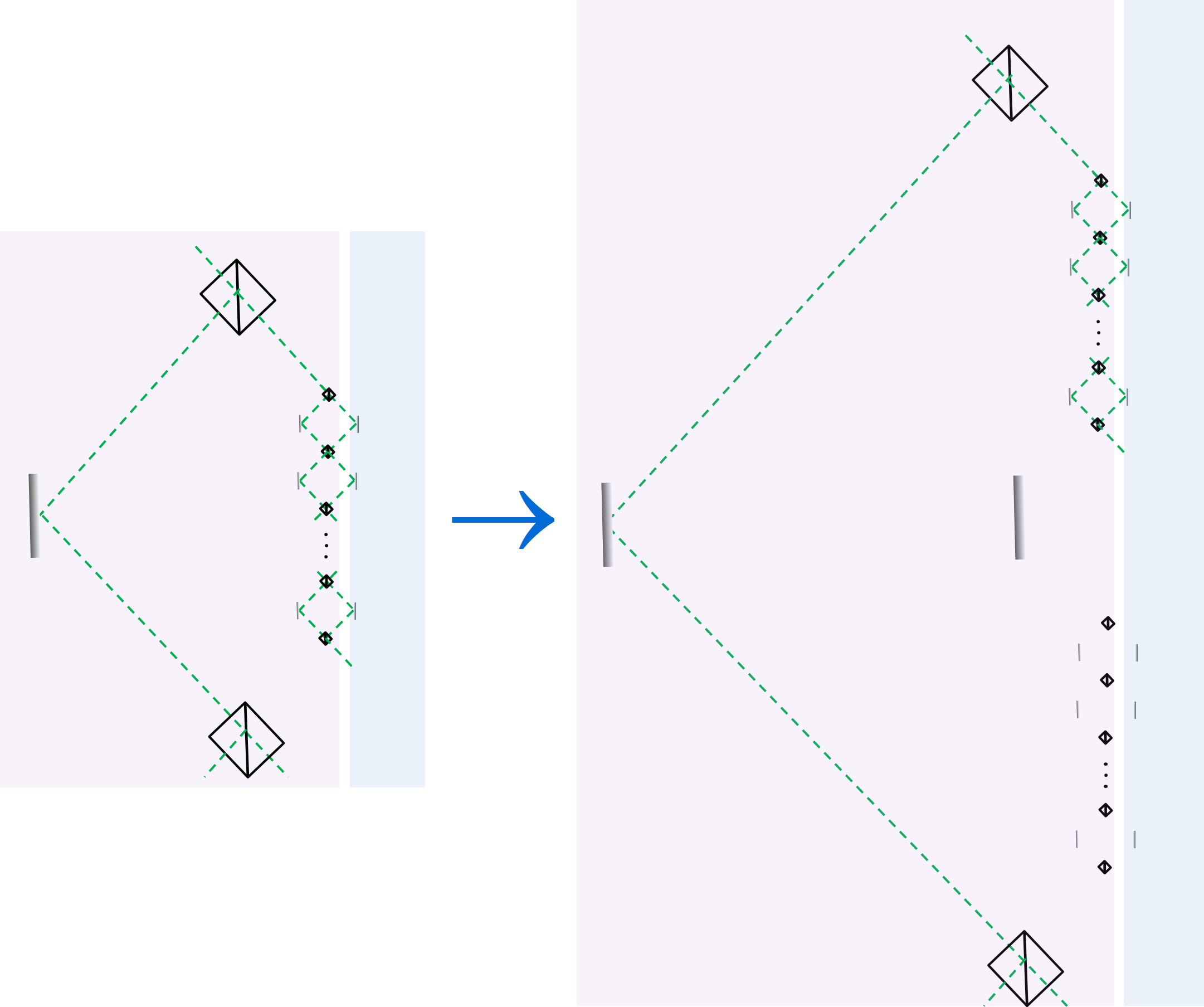}
  \caption{{\bf Modification of the protocol described in Fig.~4 which removes the weak trace. }  Each interferometer of the external chain which includes a chain of inner interferometers is replaced by the interferometer with two chains of small interferometers connected by a mirror. If there are no blocks present, the forward-evolving state of the probe does not enter the second chain, and the backward-evolving state does not enter the first chain.   }
 \end{figure}

All modified protocols are counterfactual according to the weak trace criterion. On Bob's side, there are weak traces only of second order in $\epsilon$, and making $K$ sequential MZIs instead of two, the trace can be reduced to the $K$-th order in $\epsilon$. The standard approach for calculating  the Fisher information  about parameters at Bob's site obtained by Alice yields exactly zero: No orthogonal polarization component of the probe reaches Alice.

Even a coherent model of traces, relevant for protocols as in Fig.~8, changes very little. The trace is still of second order in $\epsilon$. The Fisher information is not strictly zero, but it is proportional to $\theta^2$, so in the agreed limit of $\theta \rightarrow 0$, it vanishes again.

\section{Conclusions}

We have analyzed the counterfactuality of various communication protocols. We argued that we cannot achieve counterfactuality without postselection. It might be that in most events the protocol is counterfactual, but if we average over all events, then the rare events will introduce anomalous non-counterfactuality, so on average, whatever criterion we choose, ``counterfactual'' protocols without postselection will lose to an explicitly non-counterfactual procedure. This happens, although in a different way, for  all situations we have considered: a single path communication channel, a multiple-path communication channel, coherent and incoherent interactions.

There is full counterfactuality when the task is to communicate the presence of an opaque object in a particular place. The non-counterfactuality arises when we verify the absence of an object, or in communicating a message, when the absence of an object is one of the cases which have to be taken into account.

In our analysis we first considered a classical argument for counterfactuality according to which even an absence of an object can be found in a counterfactual way, but we argued that such a classical reasoning for quantum protocols leads to a paradox and has to be abandoned. Our derivation accords well with the results of previous analyses (for incoherent coupling) of the weak trace criterion for counterfactuality. Our results, obtained from a newly introduced criterion based on Fisher information, agree with the results of the weak trace criterion analysis.  On the one hand, it is not surprising: the weak trace is based on the local change of states of some degrees of freedom of the environment, and the  Fisher information analysis is based on the changes of some degrees of freedom of the probe, both happening due to local interactions. On the other hand, the degrees of freedom in each case are different and the mathematical analysis is also different, so the nature of our conclusions was not obvious from the beginning. Also, there were claims in the literature to the opposite \cite{A-SB}, but these were applied to analyses without postselection.

The counterfactuality analysis showed that a recently proposed modification of some counterfactual communication protocols, which is essentially applicable to all protocols, achieves the task. According to all criteria, the modified protocols are counterfactual.

The success of counterfactual communication is a surprising and paradoxical feature: we can get classical information about Bob's site, or even obtain a quantum state of an object located at Bob's site, using probes which do not leave a trace of the magnitude required by an interaction of a single probe there. This non-intuitive situation seems to require parallel worlds for a sensible local explanation  \cite{commonsense}. Counterfactual communication protocols succeed to achieve a communication task in our postselected world  due to  probes visiting Bob's site in parallel worlds.

\section*{Acknowledgements}

This work was supported in part by the Israel Science Foundation Grant No. 2064/19, the U.S.-Israel Binational Science Foundation Grant No. 735/18, the Israel Innovation Authority Grants No. 70002 and 73795, the FQXi Grant No. 224321, the Pazy Foundation, and by the Quantum Program for Early-Stage Researchers of the Israeli Council for Higher Education.

\end{document}